\documentclass[fleqn,usenatbib]{mnras}

\usepackage{newtxtext,newtxmath}

\usepackage[T1]{fontenc}
\usepackage{ae,aecompl}
\usepackage[dvipsnames]{xcolor}

\usepackage{graphicx}	
\usepackage{amsmath}	

\usepackage{amssymb}	
\usepackage[normalem]{ulem}

\title[Long-term duration changes in {\it Kepler} planets]
{Systematic search for long-term transit duration changes in {\it Kepler} transiting planets}

\author[Shahaf, Mazeh et al.]
{Sahar Shahaf,$^{1}$\thanks{E-mail:
sahar.shahaf@gmail.com}
Tsevi Mazeh,$^{1}$
Shay Zucker,$^{2}$
and
Daniel Fabrycky$^{3}$
\\
$^{1}$School of Physics and Astronomy, 
Tel Aviv University, Tel Aviv, 6997801, Israel\\
$^{2}$Porter School of the Environment and Earth Sciences, 
Tel Aviv University, Tel Aviv, 6997801, Israel\\
$^{3}$Department of Astronomy and Astrophysics, University of Chicago, Chicago, Illinois 60637\\
}

\date{Accepted 2021 May 3. Received 2021 April 14; in original form 2021 February 3}

\pubyear{2021}

\begin{document}
\label{firstpage}
\pagerange{\pageref{firstpage}--\pageref{lastpage}}
\maketitle

\begin{abstract}
Holczer, Mazeh, and collaborators (HM+16) used the {\it Kepler} four-year observations  to derive a transit-timing catalog, identifying  $260$ Kepler objects of interest (KOI) with significant transit timing variations (TTV). For KOIs with high enough SNRs, HM+16 also derived the duration and depth of their transits. In the present work, we use the duration measurements of HM+16 to systematically study the duration changes of $561$  KOIs and identify $15$ KOIs with a significant long-term linear change of transit durations and another $16$ KOIs with an intermediate significance.
We show that the observed linear trend is probably caused by a precession of the orbital plane of the transiting planet, induced in most cases by another planet. The leading term of the precession rate depends on the mass and relative inclination of the perturber, and the period ratio between the two orbits, but not on the mass and period of the transiting planet itself. 
Interestingly, our findings indicate that, as a sample, the detected time derivatives of the durations get larger as a function of the planetary orbital period, probably because short-period planetary systems display small relative inclinations. The results might indicate that short-period planets reside in relatively flattened planetary systems, suggesting these systems experienced stronger dissipation either when formed or when migrated to short orbits. This should be used as a possible clue for the formation of such systems.

\end{abstract}
%

\begin{keywords}
planets and satellites: general
\end{keywords}

\section{Introduction}
\defcitealias{holczer16}{HM+16}
Holczer, Mazeh, and collaborators \citep[][hereafter HM+16]{holczer16} analysed the {\it Kepler} data of all transiting planets known at the time, and published a catalog of transit timings for $2,599$ Kepler Objects of Interest (KOIs), identifying $260$ planet candidates that exhibited significant transit time variations (TTVs) \citep[see also][]{swift15,ofir18}. 

For $779$ KOIs with high enough signal-to-noise ratios (SNRs), \citetalias{holczer16} also derived the duration and depth of $69,914$ individual transits. This was done by adopting a general transit model for each KOI, based on stacking all its available transits, and later fitting each transit with the derived model, while varying its depth and duration. The {\it relative} change of the depth and duration and their estimated uncertainties were tabulated by \citetalias{holczer16}. 

Some previous works studied the data set of \citetalias{holczer16} \citep[e.g.,][]{kane+19} and identified additional systems with a significant TTV and also many KOIs with transit duration variations (TDV). In the present work, we return to the duration values of \citetalias{holczer16} and perform a {\it systematic} search for planets with a significant linear trend in the duration values. Following \citetalias{holczer16} we present the results in a table and accompanying figures. We refrain from analysing the depth variation, as the transit depth is more susceptible to observational artefacts like background stars and pixel shifts during the observations \citep[e.g.,][]{morton11}.

The observed linear trend of the duration is probably a manifestation of the precession of the orbital plane of the transiting planet, induced in most cases by another planet. We briefly discuss the expected timescale of the precession \citep[see][]{baileyFabrycky20} and its dependence on the parameters of the perturbing planet \citep[e.g.,][]{MiraldaEscude02}. We compare this expected dependence with our findings and suggest that our results might lead to some insights into the formation of multiple planetary systems with short-period planets \citep[e.g.,][]{architectureII14}.

Section~\ref{section:analysis} presents our analysis, and Section~\ref{section:results} lists the KOIs with a significant long-term linear change of their transit duration.
Section~\ref{section:TDVtheory} 
presents an approximated theoretical prediction for the rate of long-term linear change of transit duration induced by a perturbing inclined planet, and compares the rate of change induces by nodal to in-plane periastron precession.
Section~\ref{section:derivative} 
presents the emerging global view of the dependence of the duration derivative of the sample on the planetary period and radius. 
Section~\ref{section:discussion} discusses the TDVs found in the context of the wealth of accumulated knowledge about the TTVs of transiting planets, and considers the possible implications of our findings.

\section{Analysis}
\label{section:analysis}

\begin{figure}
\centering
{\includegraphics[width=0.5\textwidth]
{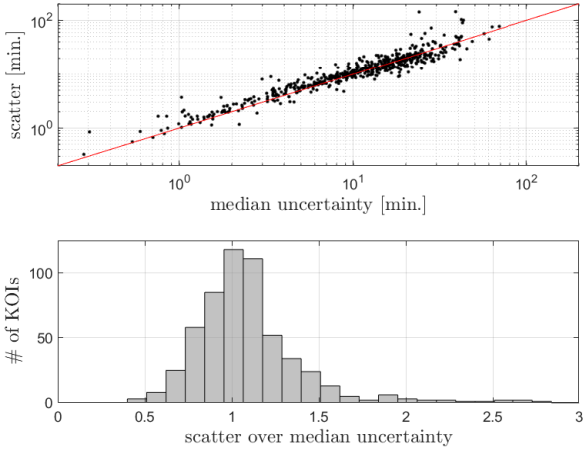}}
\caption{{\it Top}---Duration MAD of the $561$ selected KOIs as a function of their typical uncertainty. The line marks the locus where scatter equals typical uncertainty. 
{\it Bottom}---Histogram of the 
scatter-uncertainty ratio.
}
\label{figure: MADvsMed_error}
\end{figure}

The catalog of \citetalias{holczer16} includes $616$ KOIs with at least $10$ transit-duration measurements, out of which we analysed a subset of $561$ KOIs that had not been tagged as False Positives by the \textit{Kepler} team
\citep{twicken16}.
For each KOI in the restricted sample, we estimated the scatter of the obtained transit durations by their median absolute deviation \citep[MAD;][]{hampel74}, and compared it to the median of the reported duration uncertainty. We expected these two statistics to have similar values for a transiting planet without a real duration variation. The top panel of Fig.~\ref{figure: MADvsMed_error} shows that this is indeed the case for most KOIs, except for some systems that exhibit scatter values larger than expected. These planets constitute the tail of the histogram of the ratio of the two statistics, shown at the lower panel of the figure. 

We searched for a linear trend of the transit duration of each KOI by fitting a straight line to its duration measurements as a function of time.
To mitigate the effect of outliers, we performed a fit that used a robust weighted linear regression, while minimizing  the $\chi^2$ under the bisquare weight-function penalty \citep[e.g.,][]{beaton74}.\footnote{We used \texttt{Matlab}'s standard \texttt{robustfit} function, dividing the design matrix and measurements by their uncertainties.}
To each derived linear trend, we assigned a p-value that represented the probability that the fitted slope was consistent with zero, using the $t$-statistic based on the derived slope divided by its estimated uncertainty.

Transit duration may also display variations on short timescales relative to the four-year \textit{Kepler} campaign, introducing an increased correlated scatter around the long-term linear trend. To test for the presence of such correlated scatter we used the Abbe statistic  \citep[e.g.,][see Appendix A]{kendall71, lemeshko06}, $\mathcal{A}$, applied to the duration time series before and after subtracting the best-fitting linear trend.

Table~\ref{table:SampleAnalysis}, available in a machine-readable format, lists the results for all $561$ KOIs, along with the orbital period, impact parameter, and planetary radius reported in the NASA exoplanet archive KOI table.\footnote{\href{https://exoplanetarchive.ipac.caltech.edu/}{exoplanetarchive.ipac.caltech.edu}}

\begin{table*} 
\centering 
\small
\begin{tabular}{@{}crccrrrrcccc@{}} 
\hline
\hline
KOI    & ${\rm Period}\,\,\,\,$    & ${\rm Duration}$   & $b$ & 
$R_{\rm p}\,\, $ &$\sigma\,\,\,\,$  & $\Delta\,\,\,\,\,$ &     $\dot{T}_{_{\rm D}}\quad$      & p-value& $\mathcal{A}$  & $\mathcal{A}_{\rm res}$  & $N$   \\ 
        & [day] \,\,\,\,              & [hr]                &           & $[R_\oplus]$  &  [min]  & [min] &  [${\rm min/year}$]     &          &                 &      \\ 
 \hline 
$0001.01$ & $2.47061$ & $1.743$ & $0.818$ & $13$ & $0.0054$ & $0.0047$ & $-0.035\,\, (0.015)$ & $2.0E-02$ & $0.97$ & $0.99 $ & $400$ \\ 
$0002.01$ & $2.20474$ & $3.889$ & $0.224$ & $16$ & $0.0092$ & $0.0089$ & $0.014\,\, (0.022)$ & $5.2E-01$ & $1.02$ & $1.03 $ & $554$ \\ 
$0003.01$ & $4.88780$ & $2.363$ & $0.054$ & $4.8$ & $0.014$ & $0.0099$ & $-0.142\,\, (0.054)$ & $9.9E-03$ & $0.94$ & $0.97 $ & $200$ \\ 
$0005.01$ & $4.78033$ & $2.025$ & $0.952$ & $7.1$ & $0.093$ & $0.087$ & $0.21\,\, (0.31)$ & $5.0E-01$ & $0.94$ & $0.94 $ & $267$ \\ 
$0007.01$ & $3.21367$ & $3.982$ & $0.021$ & $4.1$ & $0.14$ & $0.14$ & $0.24\,\, (0.45)$ & $5.9E-01$ & $0.98$ & $0.98 $ & $312$ \\ 
$0010.01$ & $3.52250$ & $3.198$ & $0.631$ & $15$ & $0.028$ & $0.026$ & $-0.116\,\, (0.080)$ & $1.5E-01$ & $0.89$ & $0.90 $ & $370$ \\ 
$0012.01$ & $17.85522$ & $7.413$ & $0.075$ & $13$ & $0.027$ & $0.014$ & $-0.48\,\, (0.21)$ & $2.3E-02$ & $0.85$ & $0.94 $ & $69$ \\ 
$0013.01$ & $1.76359$ & $3.169$ & $0.512$ & $21$ & $0.014$ & $0.0051$ & $0.592\,\, (0.014)$ & $1.0E-10$ & $0.29$ & $1.02 $ & $679$ \\ 
$0017.01$ & $3.23470$ & $3.594$ & $0.079$ & $13$ & $0.013$ & $0.014$ & $-0.026\,\, (0.041)$ & $5.2E-01$ & $0.99$ & $0.99 $ & $320$ \\ 
$0018.01$ & $3.54847$ & $4.569$ & $0.075$ & $15$ & $0.021$ & $0.022$ & $-0.027\,\, (0.063)$ & $6.7E-01$ & $0.97$ & $0.97 $ & $357$ \\ 
\hline 
\hline 
\end{tabular}
\caption{Transit details and variance analysis of the duration measurements for the 561 KOIs. 
KOI number, period, impact parameter $b$ and planetary radius, $R_P$, are taken from the NASA exoplanet catalog (KOI table).  The duration scatter around the mean is $\sigma$,  and $\Delta$ is the median error. The derived linear trend, its uncertainty, and its p-value are given next. $\mathcal{A}$ and $\mathcal{A}_{\rm res}$ are the Abbe values of the duration measurements before and after the linear fit. $N$ is the number of transits with a duration measurement. The full table is available in the online supplementary data.}
\label{table:SampleAnalysis}
\end{table*}

\begin{figure}
\centering
{\includegraphics[width=0.5\textwidth]{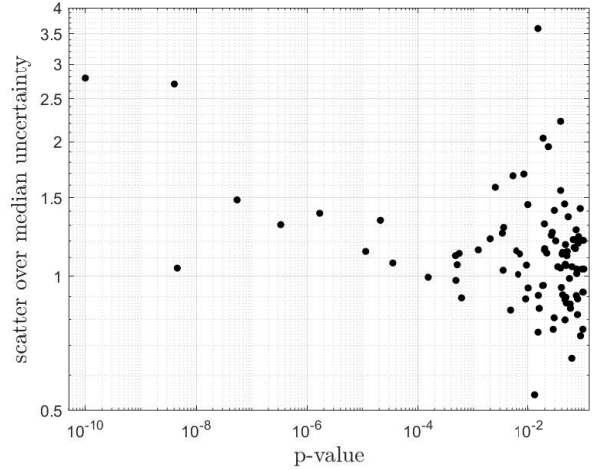}}
\caption{Scatter-uncertainty ratio vs.~p-value of the linear trend found for the $93$ KOIs with p-value lower than $0.1$.}
\label{figure: ScatterUncertaintyPval}
\end{figure}

Fig.~\ref{figure: ScatterUncertaintyPval} presents the scatter-to-uncertainty ratio
versus the derived p-values for $93$ systems with p-values lower than $0.1$. As expected, a strong correlation between the two variables can be seen in the diagram for KOIs with a p-value lower than $\sim10^{-3}$. The only obvious outlier is KOI-824.01, with a p-value smaller than $10^{-8}$ and, nevertheless, scatter almost equal to its typical uncertainty. The reason for this anomaly can be traced to a large gap in the data that separates the first six points from the rest of the measurements of this planet (See Fig.~\ref{figure: TDV signif 2}).

\section{Results}
\label{section:results}
\begin{figure}
\centering
{\includegraphics[width=0.5\textwidth]{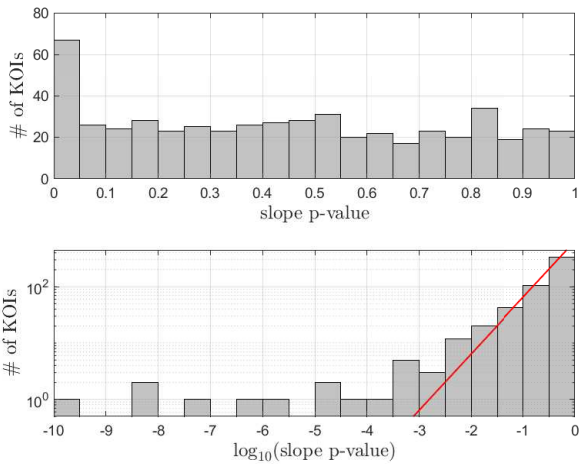}}
\caption{{\it Top}---Histogram of the p-value of the slope significance of the $561$ analysed KOIs. The p-values are uniformly distributed, except for the leftmost bin. {\it Bottom}---Histogram of the $\log$ p-values, showing the excess of targets with significant slopes. The red line represents a uniform distribution, for comparison.
}
\label{figure: pval hist}
\end{figure}

%
A histogram of the p-values of the linear slopes of all $561$ transiting planets is shown in Fig.~\ref{figure: pval hist}, displaying a large excess of planets with significant slopes. 

To exclude planets with large short-term modulation, we did not consider planets that display $\mathcal{A}<0.8$ relative to their fitted linear trend. This excluded $31$ systems from our sample, including KOI-1546.01 (see Fig.~\ref{figure: TDV signif 3} below), which deserve a special discussion, beyond the scope of this paper. We were left with a sample of $530$ planets for further study.

To identify the population of planets with a significant linear trend we applied the false-discovery rate (FDR) approach of \citet[][]{benjamini95}, designed to control the expected proportion of false discoveries. In the context of this work, false discoveries are planets with an insignificant linear trend that are falsely considered to have a significant slope, i.e., planets for which the null hypothesis of insignificant modulation was falsely rejected. The FDR procedure controls the purity of the resulting sample of discoveries. It optimizes for the fraction of false rejections of the null hypothesis within the sample of null-hypothesis rejections. The conventional, more naive, procedure of using a predetermined p-value threshold controls the probability that a specific system is not a null detection. The difference is subtle, yet the FDR approach is the preferred way to control the sample purity, as was proven by \citeauthor{benjamini95}.

\begin{figure}
\centering
{\includegraphics[width=0.5\textwidth]{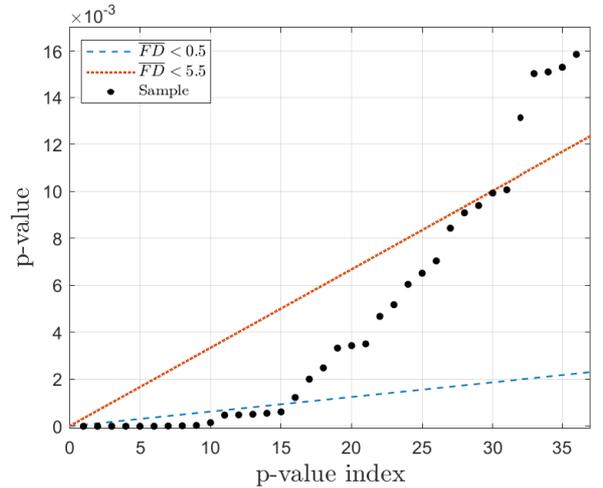}}
\caption{\citeauthor{benjamini95} FDR diagram of the TDV linear trend significance. The p-values of the first 37 planets are plotted versus their ascending-order index. The \citeauthor{benjamini95} selection criteria for the high- and intermediate-significance population are plotted as blue-dashed and dotted-red lines, respectively (see text).
}
\label{figure:FDR}
\end{figure}

The \citeauthor{benjamini95} selection procedure proceeds as follows: The $N$ objects in the complete sample are ordered according to their p-values, $\big\{\textrm{p}_{(1)},\, \textrm{p}_{(2)},\, \dotsc , \textrm{p}_{(N)}\big\}$, such that for any $i$, 
$\textrm{p}_{(i)}\leq \textrm{p}_{(i+1)}$ (see Fig~\ref{figure:FDR}).
The index $k(\alpha)$ is then defined as the largest index for which
\begin{equation}
    \textrm{p}_{(k)} \leq \frac{k}{N}\alpha \, ,
\end{equation}
%
where $\alpha\in(0,1]$ is the desired false-discovery {\it rate}, i.e., the fraction of detections which are false. \citet{benjamini95} proved that within the $k(\alpha)$ systems chosen this way, the expected number of false discoveries, $\overline{FD}$, is bounded by $\alpha k$. 

As can be seen in Fig.~\ref{figure:FDR}, there is a sub-sample of $15$ targets with 
$p(i)<0.001$, obtained with $\alpha=3.3\%$ (dashed-blue line), yielding $\overline{FR}<0.5$. This sub-sample is henceforth referred to as the sample with significant TDV. A larger sub-sample of $31$ systems with $p(i)<0.01007$, obtained with $\alpha=17.7\%$, yields $\overline{FR}<5.5$, shown by the dotted-red line in the figure. The $16$ additional planets will be referred to as planets with  an
intermediate significance slope. Details of the two samples are given in Table~\ref{table:Significant}, and the corresponding plots 
in Appendix~\ref{section: IndividualSystems}.

\begin{table}
\centering
\small
\begin{tabular}{@{}crrrrc@{}}
\hline
\hline
KOI  &  Kepler & ${\rm Period}\,\,\,\,$ & $\dot{T}_{_{\rm D}}\quad$ & p-value &  \#KOIs \\ &   & [day] \,\,\,\,           &   [${\rm min/year}$]      &           &           \\
 \hline
$0013.01$& $13$ b  & $1.76359$ & $0.592\,\, (0.014)$ & $1.0E-10$ & $1$ \\
$0103.01$&         & $14.91095$ & $4.57\,\, (0.98)$ & $1.1E-05$ & $1$ \\
$0137.02$& $18$ d  & $14.85891$ & $1.84\,\, (0.36)$ & $1.7E-06$ & $3$ \\
$0151.01$&         & $13.44718$ & $-6.4\,\, (1.8)$ & $4.8E-04$ & $1$ \\
$0209.02$& $117$ b & $18.79591$ & $3.08\,\, (0.83)$ & $4.8E-04$ & $2$ \\
$0319.01$&         & $46.15119$ & $5.0\,\, (1.3)$ & $5.5E-04$ & $1$ \\
$0377.01$&  $9$ b  & $19.27083$ & $1.57\,\, (0.28)$ & $3.4E-07$ & $3$ \\
$0377.02$&  $9$ c  & $38.90788$ & $-4.69\,\, (0.59)$ & $4.0E-09$ & $3$ \\
$0460.01$& $559$ b & $17.58752$ & $5.2\,\, (1.4)$ & $5.1E-04$ & $2$ \\
$0824.01$& $693$ b & $15.37562$ & $-6.11\,\, (0.86)$ & $4.6E-09$ & $1$ \\
$0872.01$&  $46$ b & $33.60123$ & $5.57\,\, (0.82)$ & $5.5E-08$ & $2$ \\
$0984.01$&         & $4.28747$ & $-1.4\,\, (0.32)$ & $2.1E-05$ & $1$ \\
$1320.01$& $816$ b & $10.50683$ & $-1.9\,\, (0.43)$ & $3.5E-05$ & $1$ \\
$1856.01$&         & $46.29911$ & $-14.4\,\, (3.2)$ & $1.5E-04$ & $1$ \\
$2698.01$&$1316$ b & $87.97242$ & $18.2\,\, (4.0)$ & $6.1E-04$ & $1$ \\
\hline
$0003.01$&   $3$ b & $4.88780$ & $-0.142\,\, (0.054)$ & $9.9E-03$ & $1$ \\
$0199.01$& $490$ b & $3.26870$ & $0.267\,\, (0.091)$ & $3.4E-03$ & $2$ \\
$0257.01$& $506$ b & $6.88341$ & $1.17\,\, (0.42)$ & $6.5E-03$ & $1$ \\
$0318.01$& $522$ b & $38.58337$ & $4.7\,\, (1.6)$ & $5.2E-03$ & $1$ \\
$0345.01$& $531$ b & $29.88488$ & $3.1\,\, (1.0)$ & $6.0E-03$ & $1$ \\
$0421.01$& $548$ b & $4.45419$ & $-0.3\,\, (0.12)$ & $9.4E-03$ & $1$ \\
$0484.01$& $572$ b & $17.20523$ & $-4.9\,\, (1.7)$ & $4.7E-03$ & $1$ \\
$0492.01$& $576$ b & $29.91119$ & $12.6\,\, (3.9)$ & $3.5E-03$ & $2$ \\
$0767.01$& $670$ b & $2.81650$ & $-0.24\,\, (0.077)$ & $2.0E-03$ & $1$ \\
$0841.02$&  $27$ c & $31.33046$ & $3.3\,\, (1.2)$ & $1.0E-02$ & $5$ \\ 
$0847.01$& $700$ b & $80.87235$ & $-7.9\,\, (2.5)$ & $8.4E-03$ & $1$ \\
$0883.01$&         & $2.68890$ & $-0.221\,\, (0.073)$ & $2.5E-03$ & $1$ \\
$1426.02$& $297$ c & $74.92769$ & $-2.58\,\, (0.84)$ & $9.1E-03$ & $3$ \\
$1692.01$& $314$ c & $5.96040$ & $2.37\,\, (0.71)$ & $1.2E-03$ & $2$ \\
$1801.01$& $955$ b & $14.53245$ & $-4.7\,\, (1.6)$ & $3.3E-03$ & $1$ \\
$2675.01$&$1312$ b& $5.44832$ & $1.78\,\, (0.65)$ & $7.0E-03$ & $2$ \\
\hline
\hline
\end{tabular}
    \caption{The KOIs with significant long-term linear TDV trend. The fitted duration {\it derivative} is presented along with its p-value significance. The last column gives the multiplicity of each system  according to the NASA Exoplanet Archive. }
\label{table:Significant}
\end{table}

\subsection{Known multiplicity of systems with detected long-term duration variation}

Except for KOI-13, we assume that most of the other observed duration linear trends are due to the interaction with another planet with an inclined orbit. The table includes only one pair of transiting planets from the same system --- KOI-377.01 and 0.2. They come from the {\it Kepler}-9 system \citep[e.g.,][]{kepler9-10}, the first system to display highly significant TTVs. Except for KOI-13, the two {\it Kepler}-9 planets have the most significant derivatives, of opposite signs, as expected for two interacting planets transiting opposite hemispheres of the stellar disk. In this case, we can identify the corresponding perturbing planet. 

Table~\ref{table:Significant} includes another four planets with significant derivatives from known multiple planetary systems --- KOI-137.02  \citep[Kepler-18 d;][]{cochran11,boley20,li20}, for which .01 shows insignificant long-term TDV, although the two planets do show clear anti-phased TTVs,
KOI-209.02  \citep[Kepler-117 b;][]{bruno15, boley20}, .01 show insignificant long-term TDV, but significant TTV, 
KOI-460.01 \citep[Kepler-559 b;][]{gajdos19}, 
no TDV available in \citetalias{holczer16} for .02,
and KOI-872.01 (Kepler-46 b), where a non-transiting planet was detected via TTV \citep[Kepler-46 c;][]{nesvorny12,saad17} and no TDV is available in \citetalias{holczer16} for .02 \citep[Kepler-46 d;][]{rowe14}.

There are additional six planets with an intermediate significance --- KOI-199.01 (no TDV available for .02), KOI-492.01 (no TDV available for .02), KOI-841.02 (a five transiting planet system; .01 shows anti-phased TTV, but insignificant long-term TDV), KOI-1426.02 (a three transiting planet system, locked in mean-motion resonance; .01 shows anti-phase TTV with .02, but insignificant long-term TDV), KOI-1692.01 (no TDV available for .02), and KOI-2675.01 (no TDV available for .02).

In all other $9$ high- and $10$ intermediate-significance systems, the presumed perturbing planets are not known.

In summary, for most of the planets with a detected long-term linear change of the transit duration, the perturbing planet is not known. This is either because the perturbing planet is not transiting, which probably implies a significant mutual inclination, or because its transit is too shallow to be detected, which implies a small radius (and mass).

For the planets with a known additional planet,  
the accompanying planets, except for KOI-377.01 and 0.2, do not show a significant linear slope. Either the SNR of the transits of the adjacent planets is too low, so \citetalias{holczer16} did not derive their transit duration change, or their long-term linear duration variations are too small to be significantly detected. For those systems, the upper limits of the duration derivatives might carry important constraints on their architecture. However, this is out of the scope of the present paper.


Obviously, a combination of TTVs and TDVs observations offer much better constraints on the planetary system. 
See, for example,  
\citet[][]{almenara2015} studying Kepler-117 (KOI-209, note our analysis above);
\citet[][]{freudenthal18}, \citet[][]{borsato19} and a review by \citet{ragozzine19} for Kepler-9 (KOI-377, note our analysis above);
\citet[][]{dawson20} and \citet{masuda17} for Kepler-693 (KOI-824; note our analysis above);
\citet[][]{hamann19} for K2-146; 
\citet{nesvorny13} for Kepler-88 (KOI-142);
and \citet{millsFabrycky17} for  Kepler-108 (KOI-119; too few observed transits to be included in our analysis). 
Interestingly, the latter requires a large
mutual inclination of $\sim20^{\circ}$ to account for the TTVs {\it and} the TDVs without invoking additional planets. 

In our sample, most of the $15$ planets with significant long-term duration variation show also large TTVs, probably induced by the same planet whose orbit is near
a mean motion resonance (MMR)
with the perturbed planet. 
The rest, KOI-13.01 (Kepler-13), KOI-151.01, KOI-460.01 (Kepler-559) and KOI-2698.01 (Kepler-1316), do {\it not} show significant TTVs (see the corresponding figures in the Appendix).
The duration variation of KOI-13 \citep[e.g.,][]{szabo11, herman18, szabo20} is due to interaction with misaligned stellar oblateness, so we do not expect TTVs. The orbital periods of the other three systems with no clear TTV are too long to be caused by the interaction with the stellar spin. All three do not have known additional transiting planets, but probably have additional planets with relatively large inclination and a period ratio far from commensurability. 


To discuss the results obtained here, we proceed in the next section to estimate the predicted long-term linear trend of transit duration induced by an additional external planet, based on simplifying approximations of the planetary system orbital dynamics.

\section{TDV --- Approximate Theoretical considerations}
\label{section:TDVtheory}

In most cases, the TDVs are due to nodal precession of the transiting planet orbital plane, which modifies the orbital inclination relative to our line of sight, and consequently varying the impact parameter and the length of the transit cord.
The precession is induced by the interaction of the planet's orbital momentum with an {\it inclined} spin of another body in the system; either that of the star itself, in a very few cases, like KOI-13, or that of another, adjacent planet. We proceed now to consider the timescale of the precession induced by an interaction with another planet and the resulting {\it derivative} of the transit duration. 

\subsection{Timescale of the nodal precession}

The typical precession timescale can be written, to first order, as the ratio between  the tidal torque exerted by the other planet on the transiting planet, $\tau_{tidal}$, and the angular momentum of the transiting planet, $L_{\rm p}$,
\begin{equation}
 P_{\rm prec} \sim \frac{L_{\rm p}}{\tau_{\rm tidal}}\sim \frac{M_\star}{m_{_2}}{P}\left(\frac{P_{_2}}{P}\right)^2\, ,
\end{equation}
where $M_\star$ is the stellar mass, $m_{_2}$ and $P_{_2}$ are the mass and orbital period of the perturber. 
The precession period can be of the order of $10$--$1000$ years, if, for example, $P\sim 10$ days and $M_\star/m_{_2}\sim10^3$--$10^4$.

{\it Kepler} observations span in most cases only a small section of the precession period. Therefore, the observable at hand is the osculating derivative of the duration, which is inversely proportional to the precession period. 

\subsection{The osculating {\it derivative} of the duration}

To derive the osculating {\it derivative} of the duration we note that 
the time it takes for a transiting planet in a circular orbit to cross the luminous disk of its 
stellar host, $T_{\rm_D}$, is given by
%
\begin{equation}
  \label{eq: Td}
  T_{\rm _D} \simeq \frac{P}{\pi}\frac{R_\star}{a}
  \sqrt{1-b^2} \,  ,
\end{equation}
%
where $P$ and $a$ are the orbital period and radius of the transiting planet, $R_\star$ is the stellar radius, and $b$ is the impact parameter of the planetary orbit
\citep[see, for example,][]{seager03}.
Assuming the change in transit duration is due to the change in orbital orientation, the logarithmic derivative of ${T}_{\rm _D}$ is
%
\begin{equation}
    \label{eq: TDV-impact}
    \frac{\dot{T}_{\rm _D}}{T\rm _{_D}}\simeq -\bigg(\frac{b}{1-b^2}\bigg)\dot{b} \, .
\end{equation}
%
Neglecting $r_p$, the planetary radius, relative to $R_\star$, the impact parameter is given by
%
\begin{equation}
b=({a}/{R_\star}) \cos{i_{\rm p}}\ ,
\end{equation}
where $i_{\rm p}$ is the orbital inclination of the transiting planet, defined with respect to the plane of the sky. Therefore, equation~(\ref{eq: TDV-impact}) can be written as
%
\begin{equation}
    \label{eq: TDV-impact-didt}
    \frac{\dot{T}_{\rm _D}}{T_{\rm _D}}\simeq \frac{a}{R_\star}\frac{b}{1-b^2}\frac{di_{\rm p}}{dt} \ ,
\end{equation}
%
where we use the fact that the inclination is presumably close to $90^\circ$ and therefore $\sin{i}\simeq 1$. 
Now,
%
\begin{equation}
   \frac{di_{\rm p}}{dt} \sim 
   \frac{1}{P_{\rm prec}}
   \sim
    \frac{m_2}{M_\star}\frac{1}{P}\, 
    \left(\frac{P}P_{_2}\right)^2 \, ,
 \end{equation}
%
which implies that the change in transit duration is
\begin{equation}
    \label{eq: TDV-impact-fractional}
    \frac{\dot{T}_{\rm _D}}{T\rm _{_D}}\simeq
    \eta\,\pi\,\frac{a}{R_\star}\frac{b}{1-b^2} 
    \frac{m_{_2}}{M_\star} \frac{1}{P}\, 
    \left(\frac{P}{P_2}\right)^2\, , 
\end{equation}
%
where $\eta$ is a unitless factor that originates from the details of the system's architecture.
Using equation~(\ref{eq: Td}) we finally obtain an approximate dimensionless expression for the rate of change of the transit duration,
%
\begin{equation}
\label{eq:derivative}
    \dot{T}_{\rm _D}\simeq{\eta}\frac{b}{\sqrt{1-b^2}}\bigg(\frac{P}{P_{_2}}\bigg)^2\frac{m_{_2}}{M_\star}\, .
\end{equation}


Note that we have considered only perturbing objects with circular orbits. The impact of eccentricity is expected to increase the precession rate and therefore the duration derivative by a factor of $\sim(1-e^2)^{-1.5}$ \citep[see for example][]{antognini15}, which is of order unity for mild eccentricities.

In the limit where the ratio between the planetary and stellar radii is small,  $R_{\rm p}/R_\star  \ll b$, equation~(\ref{eq: TDV-impact-fractional}) is consistent with the careful derivation of precession-induced transit duration variation given by \citet[][see equation 11 therein]{MiraldaEscude02}. 
From that equation, one can show that the geometrical factor, $\eta$, is 
\begin{equation}
\eta = 0.75\sin{2 i_{\rm rel}}\cos{\beta}\sin{\Omega_{\rm p}}\, .
\end{equation}
Here, $ i_{\rm rel}$ is the relative inclination of the two orbital planes, 
$\beta$ is the angle of the invariable plane (the plane perpendicular to the total angular momentum vector) relative to our line of sight, and $\Omega_{\rm p}$ is the node of the orbit of the transiting planet measured in the invariable plane, relative to the intersection of that plane with the plane of the sky. The angle $\Omega_{\rm p}$ completes a full revolution during the precession period. 
Through the precession the derivative oscillates between two extreme cases with $\sin \Omega_{\rm p}$ between $1$ and $-1$. For small $i_{\rm rel}$, and $i_{\rm p}$ close to $90^{\circ}$, we get $\cos \beta \simeq 1$ , and therefore 
%
\begin{equation}
- 1.5\, i_{\rm rel}\lesssim   \eta \lesssim 1.5\, i_{\rm rel} \, .
\end{equation}

To summarize, this simplistic theoretical expansion indicates that the transit-duration derivative 
is proportional to the relative angle (for small $i_{\rm rel}$) of the two planes of motion, the period ratio squared and to the mass of the perturbing planet. The derivative does {\it not} depend on the period of the perturbed planet, nor its mass.

Planetary precession may also be caused by the interaction with the stellar quadruple moment. Similarly to the development presented above, one can show that in this case the duration derivative scales like $|\dot{T}_{\rm _D}| \propto P^{-4/3}$ (see Appendix~\ref{section: quad}). This is contrary to the duration derivative due to the interaction with another planet, which depends on the ratio of the orbital periods rather than on the orbital period itself.

\subsection{Comparison to in-plane periastron precession}

We have focused on nodal precession that induces motion out of the orbital plane as the main driver of duration changes. We turn now to justify why this dynamical effect likely dominates the in-plane periastron precession.

We begin with a heuristic argument. Suppose the transiting planet has an inclination of order one radian relative to the invariable plane. Then its node would need to precess an angle $\sim R_\star/a$ to obtain a full effect on the light curve, i.e. to go from centrally transiting to non-transiting the parent star. 
Similarly, suppose the eccentricity is on the order of unity. 
To cause a central transit to go from minimum duration to maximum duration, the apse would need to precess in the plane by $\pi$ radians, changing from the periastron to the apastron pointing at the observer. On the other hand, secular torques --- either planet-planet interactions or stellar asphericity --- cause similar nodal and apsidal precession rates. Therefore, the nodal precession has the advantage over apsidal precession, by $\mathcal{O}(a/R_\star)$ in causing duration changes. 

Of course, mutual inclinations and eccentricities are much less than unity for most systems, but if they are at least on the same order of magnitude, the argument holds, as $a/R_\star$ is greater than $\sim10$ for most known exoplanets. \cite{dawson20} recently gave a numerical example to argue the same point: a planet with $a/R_\star=107.5$ and impact parameter $b=0.52$ would have equivalent duration changes via (i) an inclination change of $0.001^\circ$, (ii) a change in eccentricity from $0.5$ to $0.6$, or (iii) an in-plane precession of $35^\circ$ of an $e=0.5$ orbit. That example demonstrates that duration changes are orders of magnitude more sensitive to out-of-plane motion due to mutual inclination.

This heuristic argument justifies the focus on inclination rather than in-plane precession of the periastron, but it also guides the following full calculation.
In Appendix \ref{section: PrecessionRatio} we show that the transit duration depends most strongly on the $z$-components of the angular momentum $\mathbf{j}$ and of the eccentricity vector $\mathbf{e}$, where $z$ is the direction towards the observer. We also show by secular perturbations how these quantities depend on the properties of an additional planet. Using equations \ref{eqn:djdt}-\ref{eqn:dTde}, keeping terms up to first order in eccentricity and mutual inclination, we proceed to find the ratio, $\mathcal{R}$, of $\dot{T}_{\rm _D}$ caused by mutual inclination and by in-plane precession: 
%
\begin{equation}
\begin{aligned}
\mathcal{R} 
&= \left({\frac{\partial T_{\rm _D}}{\partial j_z}\frac{d j_z}{dt}}\right)
\bigg{/}
\left({\frac{\partial T_{\rm _D}}{\partial e_z}\frac{d e_z}{dt}}\right)\\
&=\frac{a}{R_\star}\, \frac{\sin \Delta \Omega_{\rm sky}}{e \cos \omega_{\rm sky}}\, \frac{b}{2b^2-1}\,,
\end{aligned}
\end{equation}
where $\omega_{\rm sky}$ is the argument of the periastron with respect to the plane of the sky, and $b$ is eccentricity-dependent as given in Appendix~\ref{section: PrecessionRatio}.

The main term of $\mathcal{R}$ is ${a}/{R_\star}$ --- the factor by which nodal precession is favored over in-plane precession. The next term compares the relevant component of the mutual inclination ($\Delta \Omega_{\rm sky} \equiv \Omega'_{\rm sky} - \Omega_{\rm sky}$, corresponding to the rotation of the perturbing planet's orbital plane around the line of sight, relative to that of the transiting planet) to the relevant eccentricity component of the transiting planet.  The final term depends on the impact parameter, and it is of order unity. Systems could have small mutual inclination or a tuned precession phase, such that in-plane precession causes a faster duration variation than nodal precession. However, the point is that the $a/R_\star$ term assures us that the population of duration drifts will be dominated by nodal precession. 

\section{Observed transit-duration derivatives of the sample --- a global view}
\label{section:derivative}
%
We consider now the statistical distribution of duration variation obtained here, which might reflect some features of the 3-D structure of multiple planetary systems of the {\it Kepler} sample.

For a given sample of planets with the same precession period but with different precession phases,
we expect the absolute values of the derivative to display a range from zero (for $\sin \Omega_{\rm p}=0$  or $\Delta \Omega_{\rm sky}=0$) up to a maximum value (for $|\sin \Omega_{\rm p}|=1$ or $|\Delta \Omega_{\rm sky}|=i_{\rm rel}$). As for the impact parameter factor, $\mathcal{B}=b/\sqrt{1-b^2}$, at least one can assume that the impact parameter is not close to unity, resulting in a transit too small and too shallow to be detected \citep[e.g.,][]{oshagh15}, and not close to zero, so the duration change is not observable. Noting that $\mathcal{B}\sim 1$ for $b\sim0.7$, we assume that the value of $\mathcal{B}$ is not far from unity.

One can plot the derived duration derivatives of the sample as a function of the known planetary parameters of the perturbed observed planet and look for any general trend. While $\mathcal{B}$ can be calculated from the observations and factored out, the other architectural parameters are, for the most part, unknown. This suggests that a trend should manifest as a limiting value, or an envelope, rather than a clear deterministic relation even though $\mathcal{B}$ is known. Furthermore, for a given transit duration, the impact parameter tends to be highly correlated with the fitted stellar density \citep{dawson20}, thus factoring it out may produce biased relations. Therefore, we only consider the duration derivatives, without accounting for $\mathcal{B}$.

\begin{figure*}
\centering
{\includegraphics[width=0.85\textwidth]{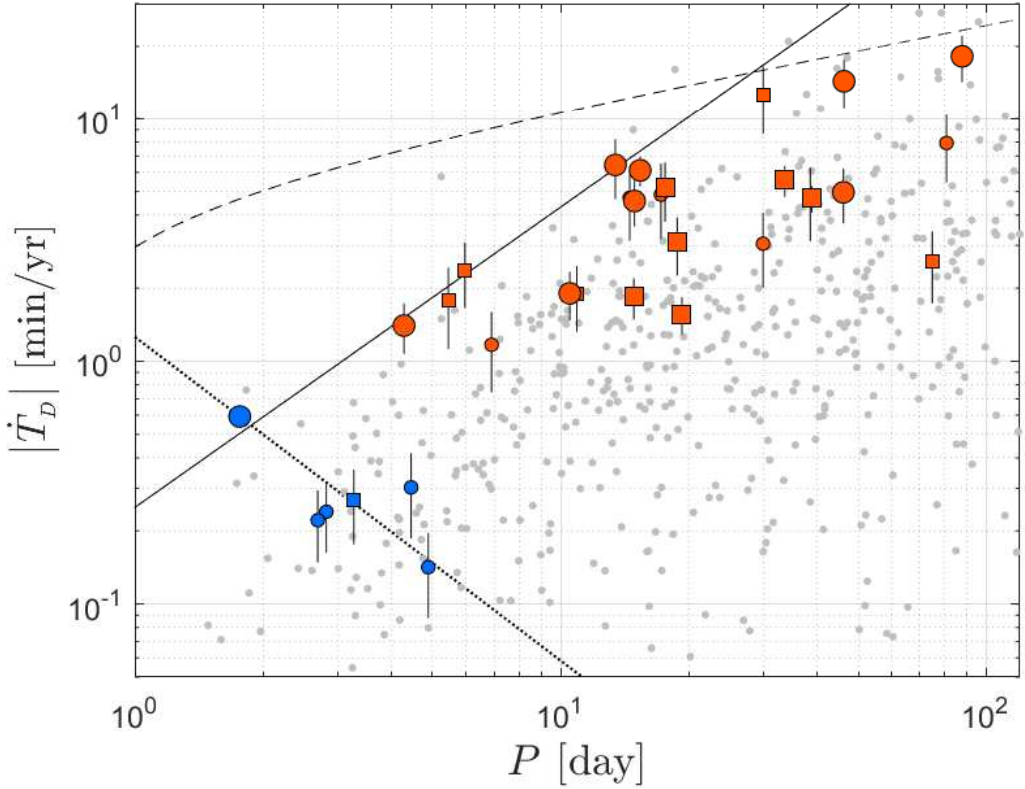}}
\caption{Absolute duration derivative vs.~the orbital period. 
Gray points represent the bulk of the sample, with slope p-values larger than $0.01007$. Small coloured circles (and squares) represent the $16$ KOIs with a slope p-value in the range of $0.001-0.01007$. The large coloured circles (and squares) represent the $15$ most significant KOIs, with p-values below $0.001$.
Squares represent systems with known additional transiting planets. Blue colour-coding marks planets for which the duration variation may have been caused by the interaction with the quadruple moment of the star (see text). The single large blue circle represents KOI-13, for which the error is too small to be plotted. 
A freely-drawn solid line represents a possible location of the upper envelope. Dashed and dotted lines represent the expected selection bias and scaling relation for interaction with the quadruple moment of the star, respectively (see text).
Some of the planets with low significance and very low derivative values are not included in the figure, due to the boundaries of the diagram, chosen for clarity.}
\label{fig: Tdor vs P}
\end{figure*}

Fig.~\ref{fig: Tdor vs P} displays the absolute values of the {\it derivatives} of the transit duration as a function of the orbital period of the transiting planet. The figure displays 
a clear paucity of large linear trends for short orbital periods, with an upper envelope, highlighted by a freely-drawn solid line of  
$\log|\dot{T}_{\rm _D}/({\rm min~yr}^{-1})| = \frac{3}{4}\log\big(\frac{P}{\rm{day}}\big) -\frac{1}{2}$.

\begin{figure*}
\centering
{\includegraphics[width=0.85\textwidth]{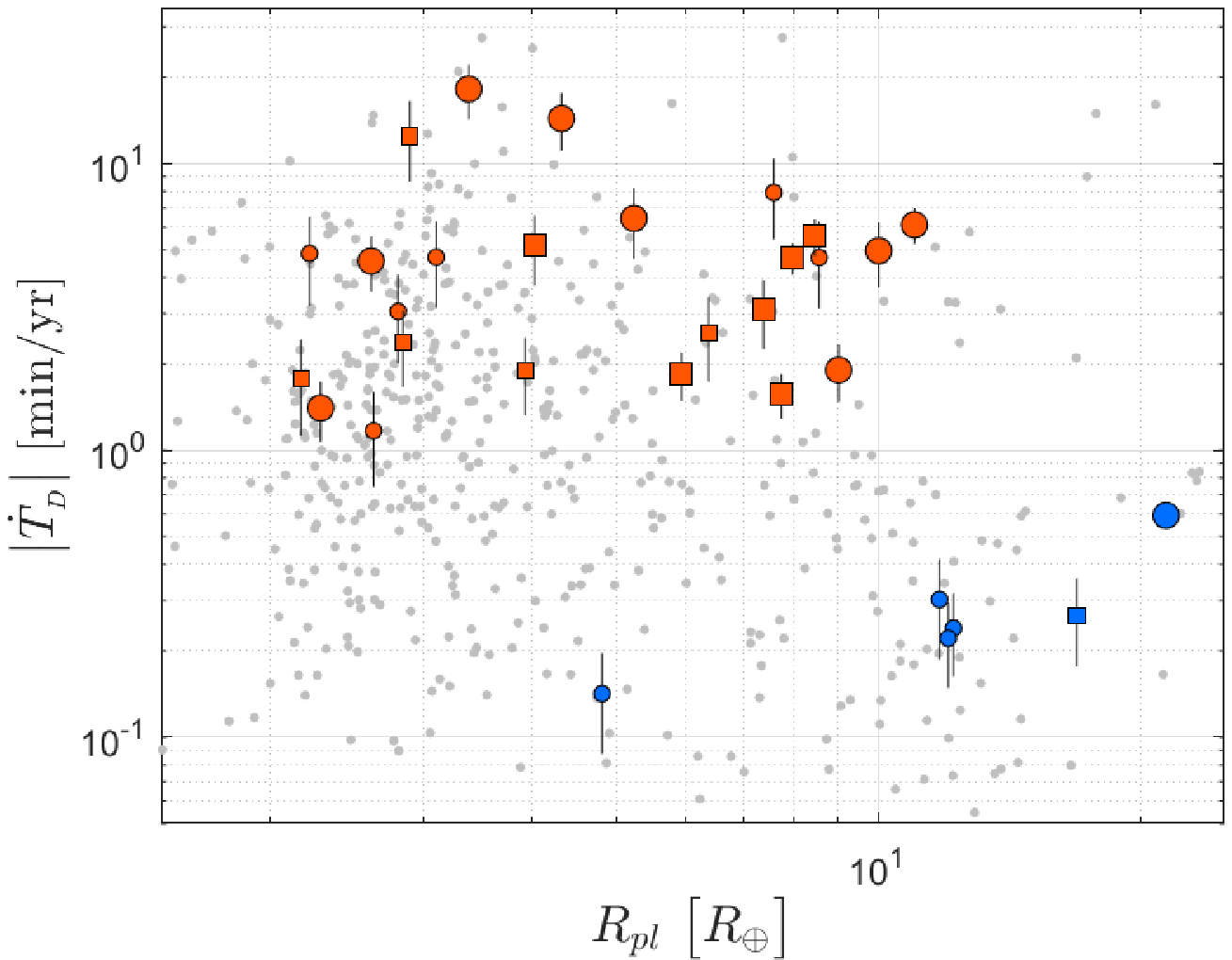}}
\caption{
Absolute duration derivative as a function of the planetary radius. 
Notation and colouring are as in Fig.~\ref{fig: Tdor vs P}.}
\label{fig: Tdor vs R}
\end{figure*}

To test the significance of this trend we divided the sample into short- and long-period sub-groups, and checked whether the short-period systems do have significantly smaller absolute duration derivatives than the long-period ones. We, somewhat arbitrarily, chose the division at $10$ days. 
To incorporate all available data, we also considered systems with insignificant fitted slopes. This is done by treating these systems as left-censored in a survival-analysis framework. A detailed description of the procedure is given in  Appendix~\ref{section: survival} (see \citealt{feigelson92} for a comprehensive review). A logrank test \citep{mantel66} rejected the null hypothesis that $|\dot{T}_{\rm _D}|$ of the two sub-groups are coming from the same distribution with a p-value of $3\times10^{-5}$. We got similar significance, between $10^{-4}$ and $10^{-5}$, for any division between $8.5$ and $12.5$ days, suggesting the trend we observe is significant.

The paucity is unlikely due to detection biases because large TDV derivatives are easier to detect than small derivatives for short- and long-period alike. 
Therefore, the absence of planets with large derivatives and short periods could either stem from a selection bias or an actual deficit of planetary systems at this region of the parameter space.

Consider the possibility that the distribution of systems in Fig.~\ref{fig: Tdor vs P} is biased by the selection criteria set by \citetalias{holczer16}. While compiling the TTV catalogue, \citetalias{holczer16} fitted the duration of individual transits only for targets with $T_{\rm _D}$ longer than $1.5$ hours and SNR better than $10$. 
As a result, for a given orbital period there is a critical value for the impact parameter above which systems were not included in the sample. 
Since $\dot{T}_{\rm _D}$ depends on the impact parameter through $\mathcal{B}$, one can use this critical value to set an upper limit on $\dot{T}_{\rm _D}$, as described in Appendix~\ref{section: selection}. The upper limit implies that with duration derivative too high, the impact parameter is necessarily high, and the duration is too small for \citetalias{holczer16} to analyse.

A dashed line in Fig.~\ref{fig: Tdor vs P} shows the expected shape of the upper envelope induced by this threshold, assuming a $15 \, \text{M}_\oplus$ outer planet with a $2:1$ period ratio, orbiting a Sun-like star. 
Equation (\ref{eq: TDV-P limit}) implies that the shape of this selection effect does not change with the period ratio or the perturber mass.
The observed upper envelope substantially deviates from the expected shape induced by the duration threshold, 
suggesting the observed paucity is due to some other effect. 

TDV caused by an interaction with the stellar quadruple moment, $J_{_2}$, is generally expected to be small compared with that induced by a nearby planet. This is the case unless the orbital period is small enough and the corresponding $J_{_2}$ is large enough, like the duration derivative of KOI-13 \citep[][]{szabo12}, which is known to be a fast-rotating A star. 

To draw our attention to this, less likely, interpretation, the dotted line in Fig.~\ref{fig: Tdor vs P} scales the TDV derivative of KOI-13 with the orbital period, assuming host stars with the same $J_{_2}$.
There are five additional systems, appearing in blue in the figure, 
with significant TDV that have derivatives close to this scaled relation --- KOIs 3.01, 199.01, 421.01, 767.01 and 883.01.
Out of the five systems, only KOI-199.01 has another reported companion, KOI-199.02, which is a known false-positive candidate. To further study this possible interpretation one should carefully estimate the $J_{_2}$ of their host stars, taking into account the relatively large value of the quadruple moment of KOI-13.

Fig.~\ref{fig: Tdor vs R} displays the absolute values of the transit derivative as a function of the radius of the transiting planet. 
It seems that the diagram might suggest that giant planets tend to show lower $|\dot{T}_{\rm _D}|$ values. 
However, the $|\dot{T}_{\rm _D}|$ distribution for planets smaller and larger than $4 \text{ R$_\oplus$}$ was compared using logrank test (see above), yielding a p-value of $0.6$. 
We, therefore, cannot conclude that the paucity of large $|\dot{T}_{\rm _D}|$ values observed for large radii has high enough  statistical significance. 

Trying to reveal what is behind the tendency displayed in Fig.~\ref{fig: Tdor vs P}, we note that the duration derivative, if resulting from the interaction between two adjacent planets, see equation~(\ref{eq:derivative}) above,
depends on:
\begin{itemize}
    \item the relative inclination between the two interacting planets, 
    \item the mass of the perturbing planet, and
    \item the period ratio of the two interacting planets.
\end{itemize}

\begin{figure*}
\centering
{\includegraphics[width=0.49\textwidth]{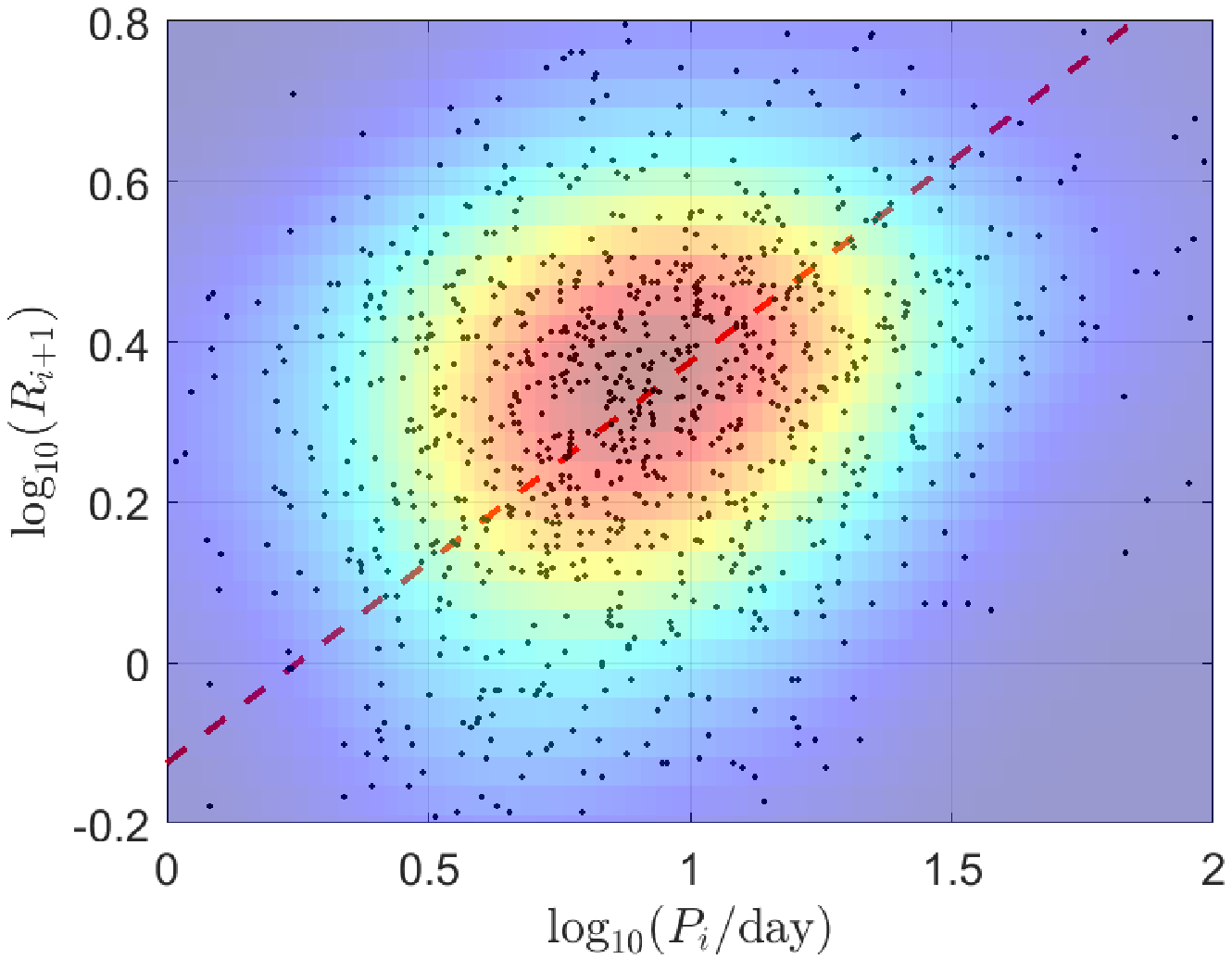}}
{\includegraphics[width=0.49\textwidth]{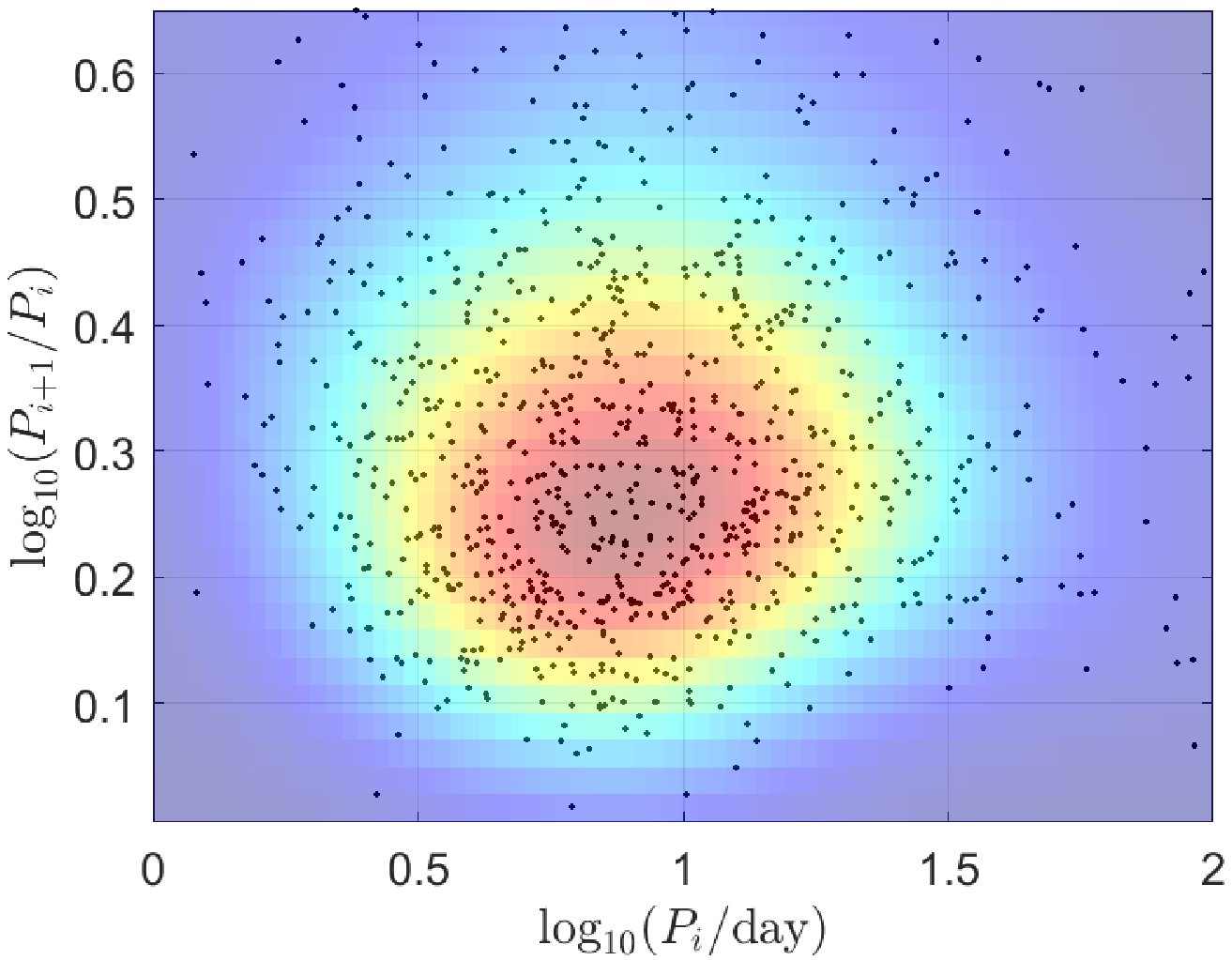}}
\caption{
{\it Left}: Planetary radius of the outer planet as a function of the orbital period of the inner planet for all pairs within the {\it Kepler} planetary multiple systems.
{\it Right}: Period ratio as a function of the orbital period of the inner planet for all pairs within the {\it Kepler} planetary multiple systems.
}
\label{fig:R_i+1 Pratio vs P}
\end{figure*}

Unfortunately, these factors are not known for most of the KOIs with a derived linear trend, as we do not observe the perturbing planet. However, we can use the whole {\it Kepler} sample of multiple systems to look at the statistical trend of the last two factors, as a function of the period in particular.
We can do that by considering all the adjacent pairs of planets in {\it known} multiple KOI systems and plotting 
(Fig.~\ref{fig:R_i+1 Pratio vs P})
the radius of the outer planet and the period ratio of the pair as a function of the period of the inner planet. 

The figure (right panel) shows 
that there is no obvious correlation between the orbital period of the inner planet and the corresponding period ratio, but there might be some correlation between the orbital period and the radius of the adjacent planet (left panel). 
This correlation stems, at least partially,
from the paucity of relatively long-period planets with {\it known} small adjacent planets. This might be an observational bias, as the longer the period the smaller the SNR of its transit. 

Furthermore, there are $135$ KOIs in the sample with an estimated duration derivative, a known outer planet candidate, and $\mathcal{A}_{\rm res} < 0.8$. As before using the \citeauthor{benjamini95} method, we examined a subset of $27$ systems with slope p-values smaller than $0.13$. This value corresponds to an expected FDR below $67\%$. The radii of the outer companions of these systems do not show any correlation with the absolute values of the detected duration derivatives. A linear regression model fitted to the logarithms of both quantities yielded an $F$-test p-value of $0.22$ and coefficient of determination, $R^2$, of ${\sim}6\%$.
We therefore conclude that the tendency of short-period planets to have smaller duration linear trends cannot be attributed to the radii of the outer planet nor the period ratios seen in Fig.~\ref{fig:R_i+1 Pratio vs P}.

\begin{figure*}
\centering
{\includegraphics[width=0.49\textwidth]{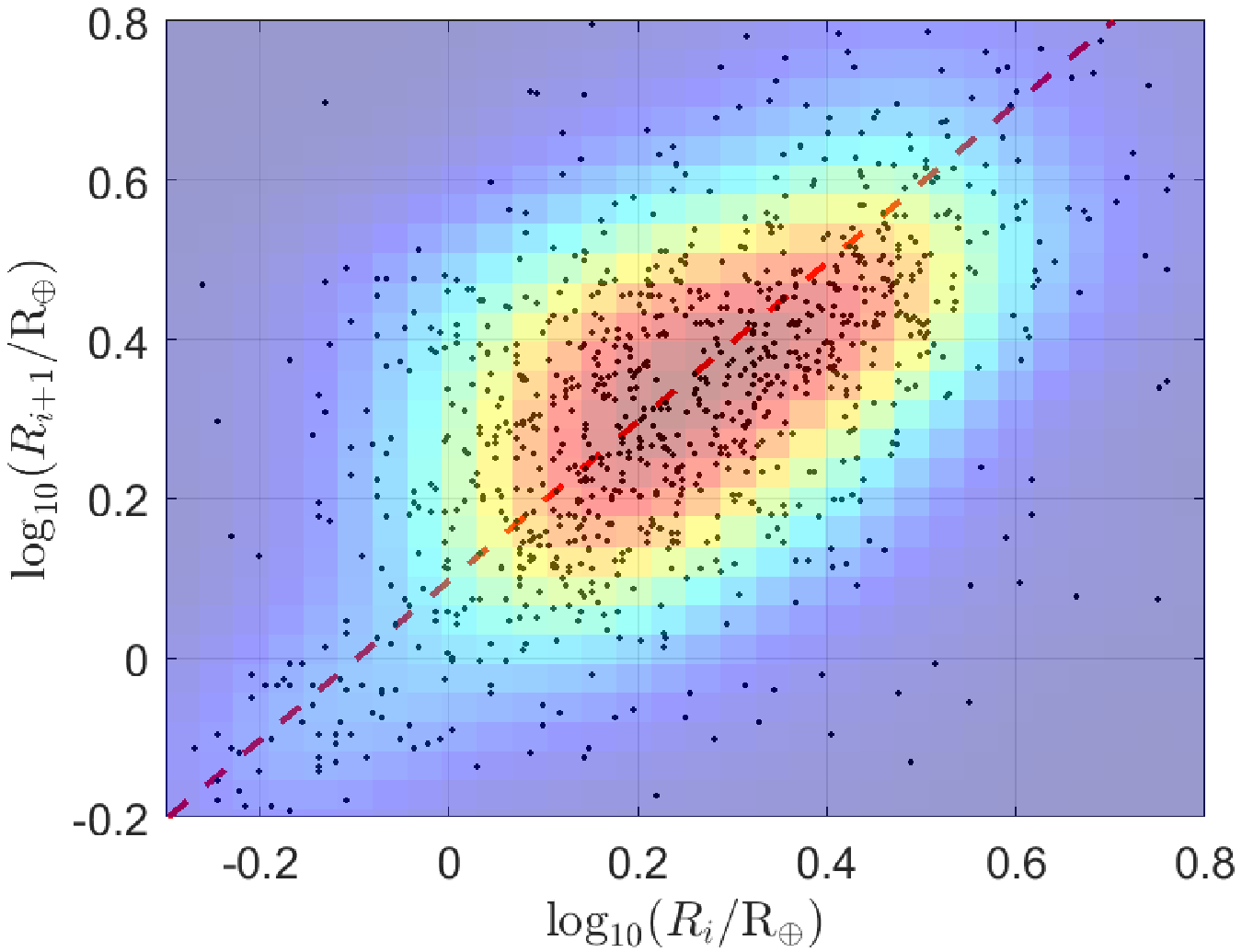}}
{\includegraphics[width=0.49\textwidth]{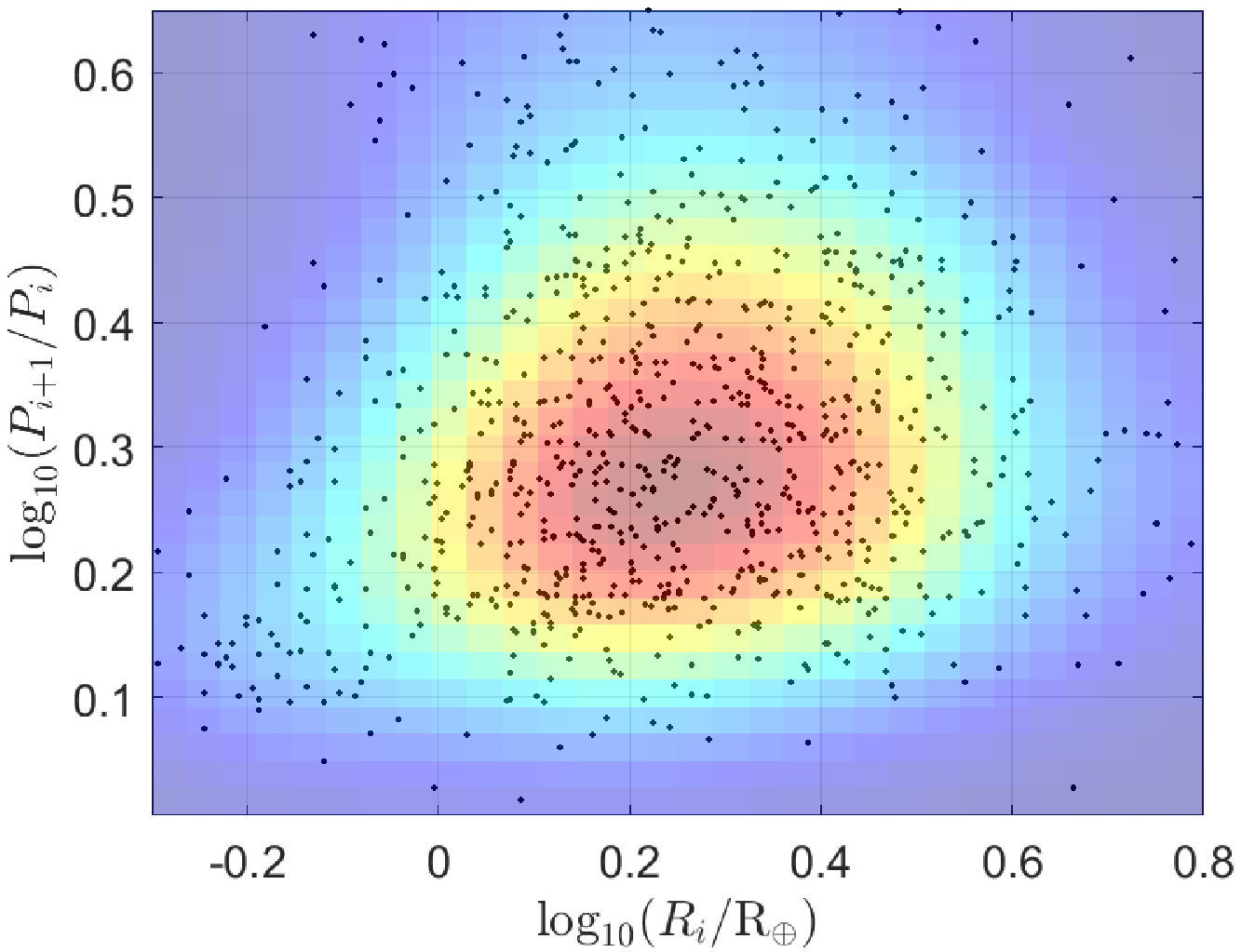}}
\caption{
{\it Left}: Planetary radius of the outer planet as a function of the planetary radius of the inner planet for all pairs within the {\it Kepler} planetary multiple systems.
{\it Right}: Period ratio as a function of the planetary radius of the inner planet for all pairs within the {\it Kepler} planetary multiple systems.
}
\label{fig:R_i+1 Pratio vs R}
\end{figure*}

For completeness, Fig.~\ref{fig:R_i+1 Pratio vs R} examines the dependence of the radius of the outer planet and the period ratio of each known pair as a function of the radius of the inner planet. 
The right panel shows 
that there is no obvious correlation between the radius of the inner planet and the corresponding period ratio, but there is a clear correlation between the radii of the two adjacent planets (left panel, also see \citealt{ford19}).

\section{Discussion and Summary}
\label{section:discussion}

The analysis presented here resulted in an easy-to-use catalogue of 561 planets with derived linear duration variation, based on the \citetalias{holczer16} catalogue. Fifteen systems showed a significant linear trend, and another 16 systems displayed a slope with an intermediate significance. The other planets did not have high enough SNR transits, and/or too small linear variation, to yield a significant linear trend.

A few systems, like KOI-142.01 \citep[Kepler-88 b;][]{nesvorny13}, show periodic short-term duration variation. These systems tend to produce an insignificant result when fitted with a linear model. Nevertheless, their variability can be identified based on their derived $\mathcal{A}$ statistic, that is provided in the catalog. A detailed characterization of these systems is beyond the scope of this work, which is focused on the long-term linear variation. 

Although the main goal of this paper is to present a systematic analysis of the long-term variability of the transit duration of a sample of KOIs with high-enough SNR transits, we wish to briefly put our results in the context of the study of the architecture and dynamics of multiple planetary systems \citep[e.g.,][]{ford19,ford20}.

The {\it Kepler} mission initiated a revolution in our knowledge of the planetary multiple systems by finding a large number of systems with more than one transiting planet. Obviously, in many multiple systems, only one planet is transiting, and therefore the multiplicity of the system remain unknown \citep[e.g.,][]{lissauer11,architectureII14}.  
The mission also opened a window that allowed the study of the dynamical interaction within multiple planetary systems. This was done in the last years by observing the TTVs \citep[e.g.,][]{agol05,holman05} of many transiting planets. See for example, a review by \citet[][]{AgolFabrycky18}.

The main effect can be quite strong for two planets with orbits near MMR.
Sometimes, another, much smaller, effect can be detected, with a substantially shorter period, termed by some studies as 'chopping' \citep[e.g.,][]{holman10,nesvorny14,deck15}. Observations of the two effects in the same system can be a powerful tool for the derivation of the parameters of the interacting planets \citep[e.g.,][]{agol21}.

The TTV became an important tool when the 4-year precise photometry of the {\it Kepler} space mission became available.
The observed TTVs have been used by many studies to infer crucial parameters of the interacting planets, like masses and relative inclinations \citep[e.g.,][]{lithwick12,hadden16,jontof16,MacDonald16, MillsMazeh17}, and even the orbital period of non-transiting planets \citep[e.g.,][]{ballard11,nesvorny12}.
 
As the TTV results from velocity change of the transiting planet along its orbit, one could expect to observe a corresponding change of the transit {\it duration}
\citep[e.g.,][]{nesvorny13}.
However, we have shown that most of the observed TDVs are due to the precession of their orbital planes, induced by another planet with an {\it inclined} orbit. 
As such, the TDVs open up another opportunity of studying the configuration and mutual inclinations of the planetary systems, 
clearing another avenue for the study of planetary multiple systems,  their 3-D structure in particular.
In a dialectic way, the more inclined the orbit of the perturbing planet is the larger the duration variation is; but, on the other hand, the chances of observing the transits of the perturbing planet are smaller. This is why we know so little about the multiplicity of some of the systems with detected duration variation.

The best we can do is to consider the period of the precession of the transiting planet, which is determined by the parameters of the two planetary orbits. However, the precession period is long relative to the 4-year time span of {\it Kepler} observations, longer than the timescale of the TTV modulation, observed in many of the systems considered here.
Therefore, even the precession period is not accessible in most cases. One can only estimate the osculating rate of change of the duration transit for planets with high enough SNR. To make things even worse, as can be seen in equation~(\ref{eq:derivative}), the linear trend of the duration change depends on two additional accidental variables --- the impact parameter and the precession phase at the time of the observations. 

Instead, we considered the absolute values of the {\it derivatives} of the transit duration of the sample of planets we had at hand as a function of the orbital period and the radius of the transiting planet. The duration derivative as a function of the orbital period displayed an upper envelope, with a clear paucity of large linear trends for short periods. 
This paucity is unlikely due to observational biases because large TDV derivatives are easier to detect than small derivatives for short- and long-period alike.

The duration derivative was not expected to depend on the period of the transiting planet. Therefore, the trend we observe is probably due to some statistical features of the architecture of planetary multiple systems. 
Trying to search for reason for this trend, we examined the correlation between the period of the inner planet and the radii of the outer planet and the period ratio, for all known pairs of planets in the {\it Kepler} multiple systems.
We found that the period ratio does not depend on the period of the inner planet, while the radius of the external planet is slightly correlated with the orbital period of the inner planet, although this can be the result of an observational bias.
As shown, this correlation is too weak to explain the upper limit we clearly see in our sample, for which planets with shorter periods tend to have smaller duration linear trends. 
Therefore, the upper envelope we observe might indicate that short-period planets tend to have smaller relative inclinations relative to their adjacent planets.

At face value, this indirect evidence is in contrast with the results of \citet[][]{dai-winn18}, who showed some direct evidence that the mutual inclination is ``larger when the inner orbit is smaller, a trend that does not appear to be a selection effect". However, their study is focused on systems for which the period of the inner planet is smaller than $\sim1.3$ days, while our sample extends to periods of $\sim 80$ days. 

For completeness, we analysed the same sample of pairs of planets in the {\it Kepler} multiple systems and found no correlation between the radii of the inner planets and the period ratios of the corresponding pairs, while large planets show a tendency to have larger adjacent planets, similar to the results of  
\citet{lissauer11, ciardi13, millholland17, Weiss18}; and \citet{ford19}. 

In summary, our analysis might tentatively suggest that the smaller duration derivatives of short-period planets might be due to smaller relative inclination. This presents probable indirect evidence that short-period planets tend to reside in more
aligned systems; their 3-D architecture is flatter than the long-period ones.

The observational and theoretical aspects of mutual inclinations of multiple planetary systems have been intensively discussed in the literature \citep[e.g.,][]{fang12, tremaine-dong12, fabrycky14, hansen15, spalding16, steffen16,
rodriguez18, zhu18, rodet-lai20}; see \citet[][]{dai-winn18} for a thorough summary of the different approaches. 
In particular, large mutual inclinations might hint at a dynamical excitation, while small relative angles could suggest
that the dissipation in the protoplanetary disk is responsible for making systems flat, and perhaps for re-flattening them after mergers but before the disk dispersal
\citep[e.g.,][and references therein]{becker-fabrycky20, spalding20}.
The possible indirect evidence presented here might serve as another tier in the discussion that might lead to a better understanding of planet formation.



\section*{Data availability}

The results of this study are available in the article and in its online supplementary material. The data underlying the analysis was taken from the \citetalias{holczer16} catalogue, which is available on line.\footnote{\href{ftp://wise-ftp.tau.ac.il/pub/tauttv/TTV/ver_112}{ftp://wise-ftp.tau.ac.il/pub/tauttv/TTV/ver{\char`_}112}}

\section*{Acknowledgements}
Part of this work was done when the authors participated in summer 2019 in a program on ``Big Data and Planets'' in the Israel Institute for Advanced Studies of Jerusalem (IIAS). We deeply thank the director, management and staff of the IIAS for creating a wonderful environment that fostered a thorough study of the subject that eventually led to this publication.

We deeply thank the referee, David Nesvorn\'{y}, for his helpful and insightful comments.
We thank Daniel Yekutieli for introducing to us the Benjamini-Hochberg method of False Discovery Rate.
We thank Amitay Sussholz for his significant contribution during the early stages of the work.
The research was supported by Grant No. 2016069
of the United States-Israel Binational Science Foundation
(BSF). It has made use of the NASA Exoplanet Archive, which is operated by the California Institute of Technology, under contract with the National Aeronautics and Space Administration under the Exoplanet Exploration Program.


\bibliographystyle{mnras}
\bibliography{Duration_variation_arXiv}

\appendix
\section{The Abbe statistic}

Consider a sequence of measurements, or residuals of measurements with regard to some model, 
$\{y_i;i=1,n\}$,
with an average of zero and a typical scatter $\epsilon=\sqrt{\sum{y_i^2}}/n$.
The Abbe statistic uses the series of consecutive differences of the series, $\Delta_i = y_{i+1} - y_{i}$, and is based on the ratio of the scatter of this difference series to the scatter of the original series:
\begin{equation}
    \mathcal{A}\equiv\frac{1}{2}\frac{1}{(n-1)} \bigg(\,{\sum_{i=0}^{n-1}{\Delta_i^2}}\bigg) \bigg/
    \epsilon \, .
\end{equation}

For an uncorrelated series of $y_i$'s, 
one may expect that a typical difference between any two points, in particular two consecutive points, $y_{i+1}-y_i$, is the quadratic sum of the characteristic scatter of the series, i.e., 
$\overline{\Delta}\approx\sqrt{2}\ \epsilon$, thus $\mathcal{A}\approx1$.

On the other hand, for a highly correlated series, one can consider two limiting cases:
\begin{itemize}
\item
Most of the pairs of consecutive measurements are close to one another, in terms of the typical scatter of the series,
$\overline{\Delta^2}\ll \epsilon^2$. In such a case $\mathcal{A}\approx0$. 
\item
For most of the pairs of consecutive measurements
$y_{i+1}\approx -y_i$, and therefore
for all measurements $|y_i|\approx\epsilon$, which means
$\overline{\Delta} \approx 2\epsilon$,
which yields $\mathcal{A}\approx 2$.
\end{itemize}

The ability of $\mathcal{A}$ to detect deviation from independence is used in our analysis, by calculating $\mathcal{A}$ before and after the fit. For example, had a significant linear trend been present in the fitted data without any additional systematic contributions, one may expect the raw data to have $\mathcal{A}$ substantially smaller than unity, and the residuals to show $\mathcal{A}\approx{1}$. These quantities were calculated and appear on Table \ref{table:SampleAnalysis}.

\section{Duration derivatives due to the two precessions}
\label{section: PrecessionRatio}

We discuss here the nodal and apse precessions induced by an external planet and the duration variation they impose on the transiting planet. We follow  \cite{2014Boue} and use two vectors that characterize the planetary orbit --- $\mathbf{e}$ and $\mathbf{j}$, where the former is the eccentricity vector, pointing towards the periastron of the transiting planet and having magnitude $e$, and the latter is the angular momentum vector, orthogonal to the orbital plane and having magnitude $\sqrt{1-e^2}$. 
The advantage of using these variables is that we may express their secular changes in an invariant way; for instance, they depend only on the mutual inclination of the orbital planes of the two planets and not on the inclinations relative to a reference plane. However, we may evaluate the results in the observer's reference frame, in which the z-axis points at the observer and the transiting planet's ascending node is on the x-axis: 
\begin{align}
    \mathbf{e} &= e \begin{pmatrix} \cos \omega \\ \sin \omega \cos i_p \\  \sin \omega \sin i_p  \end{pmatrix},\\ 
    \mathbf{j} &= \sqrt{1-e^2} \begin{pmatrix} 0 \\ -\sin i_p  \\ \cos i_p \end{pmatrix}, 
    \ \mathbf{j'} = \sqrt{1-e'^2} \begin{pmatrix} \sin i_p' \sin \Omega' \\ -\sin i_p' \cos \Omega' \\ \cos i_p' \end{pmatrix}\, . 
\end{align}  
Here we also have listed the primed coordinates, corresponding to the perturber, assumed to be on an external orbit. To emphasize that the argument of periastron $\omega$ and longitude of the node $\Omega$ are in this observer frame, we append the subscript ``sky'' when using them in the main text. The transit duration may be written in terms of these variables: 
\begin{equation}
\begin{aligned}
    T_{\rm _D} &= \frac{P R_\star}{\pi a} \frac{\sqrt{1-e^2} \sqrt{1-b^2}}{1+e \sin \omega}\\
    &=\frac{P R_\star}{\pi a} \frac{(\mathbf{j}\cdot\mathbf{j})^{1/2}}{1+(e_y^2+e_z^2)^{1/2}} \sqrt{1-b^2},
\end{aligned}
\end{equation}
where the impact parameter is 
\begin{equation}
    b = \frac{a}{R_\star} \frac{1-e^2}{1+e \sin \omega}  \cos i = \frac{ (\mathbf{j}\cdot\mathbf{j})^{1/2} j_z}{1+(e_y^2+e_z^2)^{1/2}}
\end{equation}

We will only pursue the calculation to first order in eccentricity $e$, in mutual inclination $i_{\rm rel}$, and in the tilt between the invariable plane and the line of sight $\beta$, and to lowest order in semi-major axis ratio. We may thus reduce this expression to: 
\begin{equation}
    T_{\rm _D} \approx \frac{P R_\star}{\pi a} (1-e_z) \bigg{(}1-(a/R_\star)^2 j_z^2(1-2e_z)\bigg{)}^{1/2},
\end{equation}
i.e., a function of only one of the eccentricity components and one of the inclination components. We may write its time derivative by applying the chain rule with respect to each of these: 
\begin{equation}
    \frac{d T_{\rm _D}}{dt} = \frac{\partial T_{\rm _D}}{\partial j_z} \frac{d j_z}{dt} + \frac{\partial T_{\rm _D}}{\partial e_z} \frac{d e_z}{dt}. \label{eqn:chain}
\end{equation}

The time dependence of the components is computed via Hamilton's equations \citep[equation 14]{2014Boue}: 
\begin{align}
\frac{d j_z}{dt} &= -\frac{1}{\Lambda} \bigg{(} 
  j_x \frac{\partial \bar{H}}{\partial j_y} 
- j_y \frac{\partial \bar{H}}{\partial j_x} 
+ e_x \frac{\partial \bar{H}}{\partial e_y} 
- e_y \frac{\partial \bar{H}}{\partial e_x} \bigg{)}\,,\\
\frac{d e_z}{dt} &= -\frac{1}{\Lambda} \bigg{(} 
  e_x \frac{\partial \bar{H}}{\partial j_y} 
- e_y \frac{\partial \bar{H}}{\partial j_x} 
+ j_x \frac{\partial \bar{H}}{\partial e_y} 
- j_y \frac{\partial \bar{H}}{\partial e_x} \bigg{)}\,,
\end{align}
with 
\begin{equation}
    \bar{H} = - \frac{G m m'}{a'} \bigg{(}c_1 
    + c_2 (e^2+e'^2 + \mathbf{j}\cdot\mathbf{j'}-1) 
    + c_3 (\mathbf{e}\cdot\mathbf{e'}) 
    + \mathcal{O}(e^4, i_{\rm rel}^4, e^2 i_{\rm rel}^2)\bigg{)}\,,
\end{equation}
and $c_2 = 1/2 b_{1/2}^{(0)}$ is a Laplace coefficient and $\Lambda$ is the angular momentum of a circular orbit.  We neglect the $c_3$ term as it is a higher-order Laplace coefficient (proportional to one more power in $\alpha= a/a'$ than $c_2$), and $c_1$ is a constant that does not effect the equations of motion. 

Keeping terms of the lowest order in $e$, $e'$, ($i-\pi/2$), ($i'-\pi/2$), and $\Omega'$, we find for the four terms in equation~\ref{eqn:chain}: 
\begin{align}
\frac{d j_z}{dt} &= (m'/M_\star) (2 \pi/P) \alpha c_2 \sin \Omega',  \label{eqn:djdt} \\
\frac{d e_z}{dt} &= - (m'/M_\star) (2 \pi/P) \alpha c_2 e \cos \omega ,\\
\frac{\partial T_{\rm _D}}{\partial j_z} &= - \frac{P}{\pi} \frac{b}{\sqrt{1-b^2}} , \\
\frac{\partial T_{\rm _D}}{\partial e_z} &= \frac{P R_\star}{\pi a} \frac{2b^2-1}{\sqrt{1-b^2}}. \label{eqn:dTde}
\end{align}

\section{Survival analysis of left-censored data}
\label{section: survival}
The fitted transit duration derivative is considered significant if the null hypothesis, stating that $\dot{T}_{\rm _D} = 0$, is rejected. While many of the estimated slopes are not significant enough to be considered detections, an upper limit can be set based on their estimated uncertainty, $|\dot{T}_{\rm_D}|+2\Delta \dot{T}_{\rm _D}$. In survival-analysis terminology, these targets are referred to as left-censored \citep[e.g.,][]{feigelson92}. Statistical arguments regarding the overall behaviour of the sample may be biased if the left-censored targets are discarded. 

The survival function, $S(t) \equiv \textrm{Pr}\big(|\dot{T}_{\rm _D}| > t \big)$, can be used to compare the observed distribution of different populations in the presence of censored data. It is often derived using the Kaplan-Meier estimator \citep{kaplan53}, originally designed for right-censored data. While the most common applications of survival analysis consider right-censoring, methods designed to address left- and interval-censored data also exists \citep[e.g.,][]{turnbull76, feigelson92,gomez92, gomez94}. 

The survival analysis in this work was done using the \texttt{lifelines}\footnote{\texttt{lifelines} survival analysis library documentation: \href{https://lifelines.readthedocs.io/en/latest/}{lifelines.readthedocs.io}.} package \citep[][]{pilon21}. For simplicity, we  used the scale-reversing procedure as described by \citet{gillespie2010}. First, a number larger than the maximal $|\dot{T}_{\rm _D}|$ in the sample was chosen and the measured data were subtracted from it, yielding a right-censored data set of positive values. Then, the population was divided into comparison groups, for example, short- and long-period planets.
For each group, the survival function was derived using the Kaplan-Meier estimator. Finally, the logrank test \citep[][]{mantel66} was used to test the null hypothesis that the two survival functions are identical. 

\section{Selection criterion of HM+16}
\label{section: selection}

The transit duration of a planet is a decreasing function of the impact parameter, for a given stellar radius and an orbital period. 
Therefore, for a duration catalogue like \citetalias{holczer16} for which only durations longer than $1.5$ hours were included, 
there is an inherent observational bias of having only small impact parameters for a given orbital period.
The limit for the impact parameter, $b$, can be written as 
%
\begin{equation}
    \label{eq: bcrit}
    \sqrt{1-b^2} \simeq 0.83 \bigg(\frac{P}{{\rm day}}\bigg)^{-\frac{1}{3}}\, ,
\end{equation}
where we have assumed a Sun-like stellar host.
This means, for example, that for systems with a $1$ day orbital period, only targets with $b\lesssim 0.55$ will be included in the sample.

Using equation~(\ref{eq:derivative}), and 
setting $\eta$ to its maximal value of $0.75$
and the impact parameter to its limiting value according to equation~(\ref{eq: bcrit}), we get that 
\begin{equation}
    \label{eq: TDV-P limit}
    |\dot{T}_{\rm _D}| \lesssim 1.4\, \bigg(\frac{m_{_2}}{\textsc{m}_\oplus}\bigg)
    \bigg(\frac{P}{P_{_2}}\bigg)^2
    \bigg(\frac{P}{\text{day}}\bigg)^{\frac{1}{3}}
    \sqrt{
    1 - 0.69\bigg(\frac{P}{\text{day}}\bigg)^{-\frac{2}{3}} 
    }
    \quad \frac{\text{min}}{\text{year}}\, .
\end{equation}

\section{Interaction with the stellar quadruple moment}
\label{section: quad}
Studies of KOI-13 \citep[Kepler-13;][]{szabo12} demonstrated that transit duration variation may also be caused by the interaction of a transiting planet with the quadruple moment of its host star.
The precession timescale for such an interaction is given by the ratio of the angular momentum of the transiting planet, $L_{\rm p}$, and the torque exerted on it by the star, $\tau_{\rm q}$.
Following \citet{MiraldaEscude02}, the precession timescale is given by
\begin{equation}
P_{\rm prec}\propto
\frac{L_{\rm p}}{\tau_{\rm q}} \sim   \frac{1}{J_{_2}} \bigg(\frac{a}{R_\star}\bigg)^2 {P} \,,
\end{equation}
where $J_{_2}$ is the quadruple moment of the star and $a$ is the orbital separation.

We assume, in analogy to the derivation of equation~(\ref{eq: TDV-impact-didt}), that the time derivative of the orbital inclination ${d i}/{dt}$ is inversely proportional to ${P_{\rm prec}}$.  
Following the derivation in Section~\ref{section:TDVtheory}, one obtains that $ |\dot{T}_{\rm _D}| \propto a^{-2}$,
hence,
\begin{equation}
    |\dot{T}_{\rm _D}| \propto P^{-\frac{4}{3}}.
\end{equation}

\section{Plots of individual systems }
\label{section: IndividualSystems}
\begin{figure*}
\centering
{\includegraphics[width=0.46\textwidth]{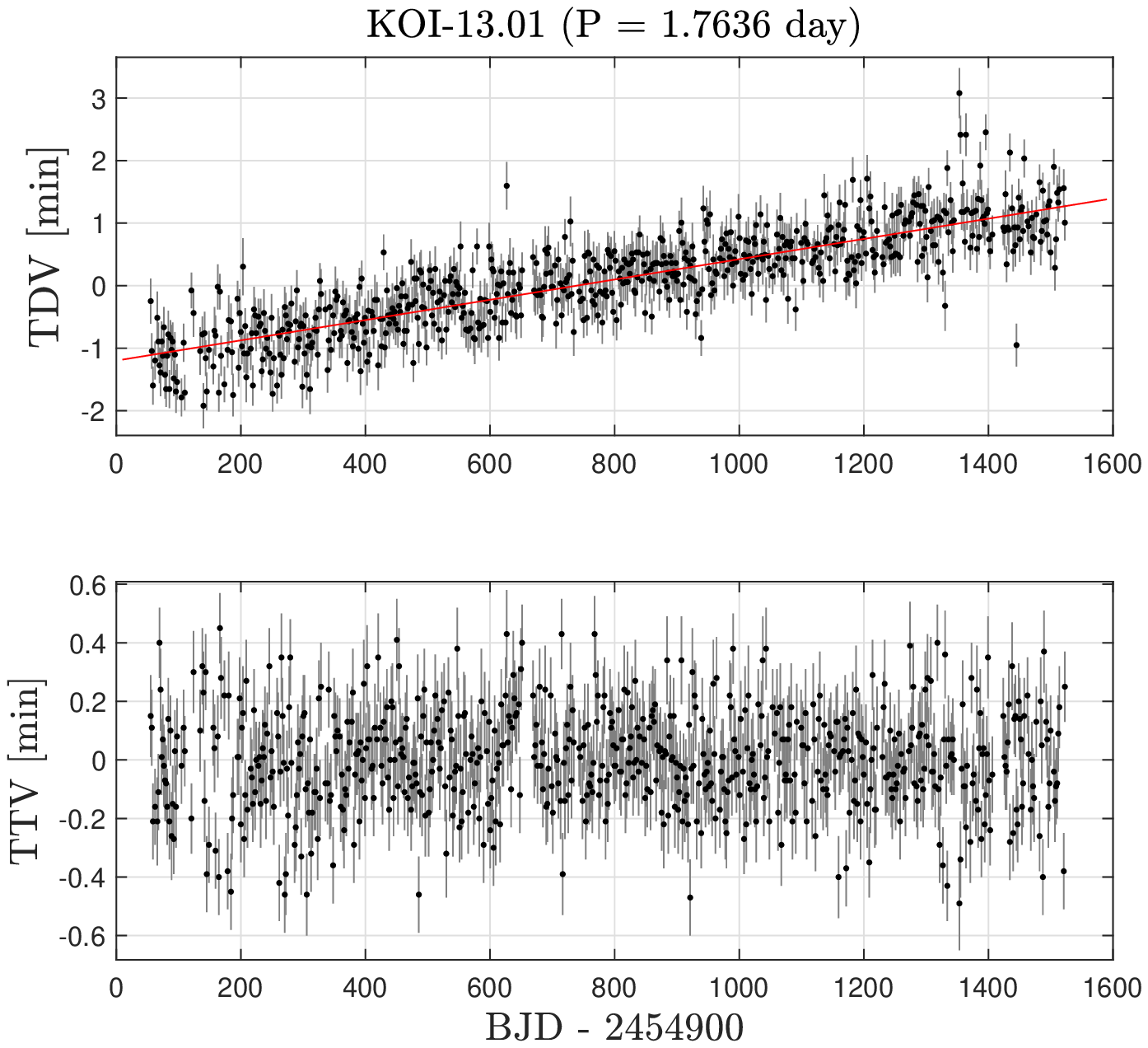}} \hspace{1mm}
{\includegraphics[width=0.46\textwidth]{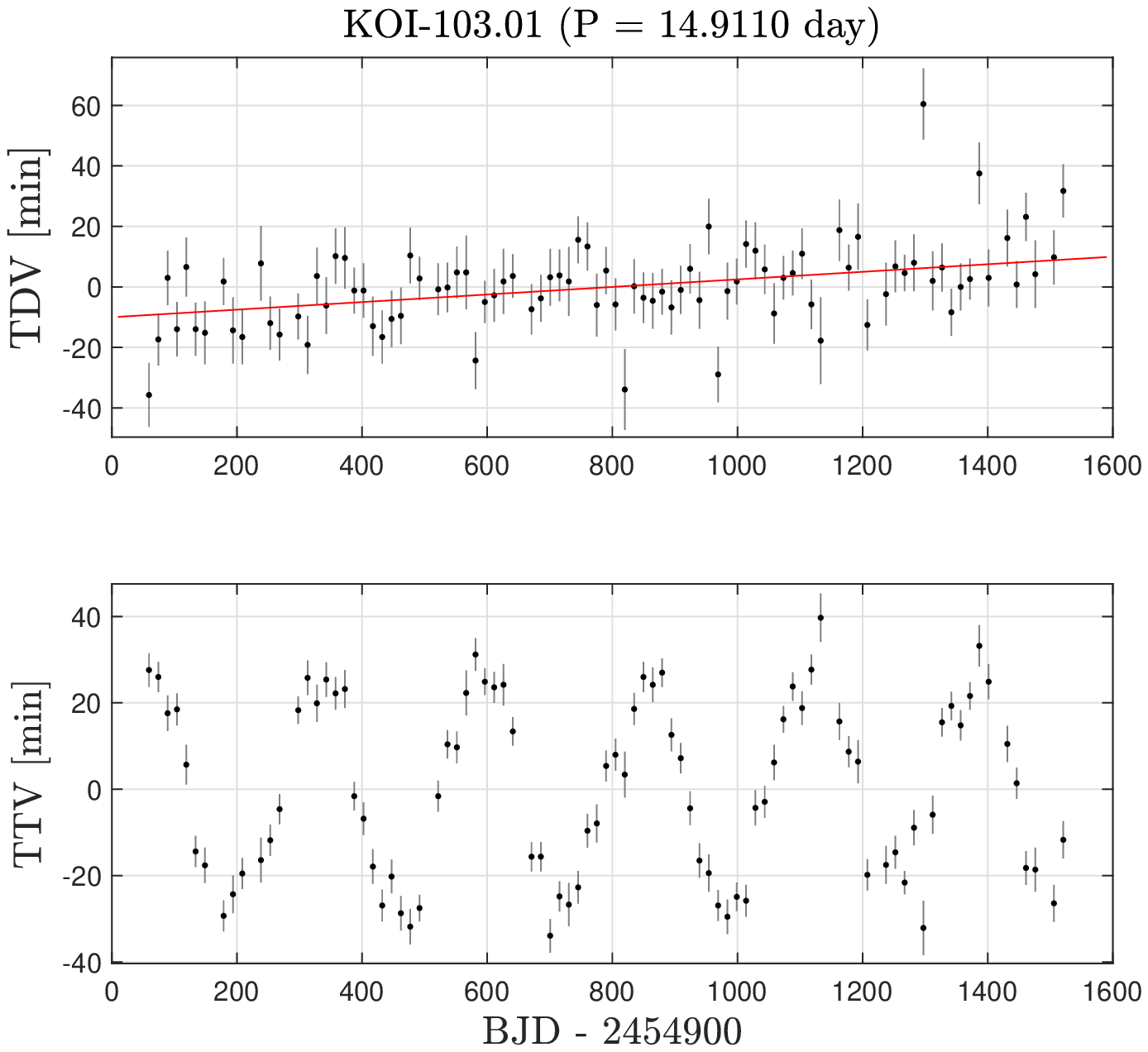}}\\ \vspace{5mm}
{\includegraphics[width=0.46\textwidth]{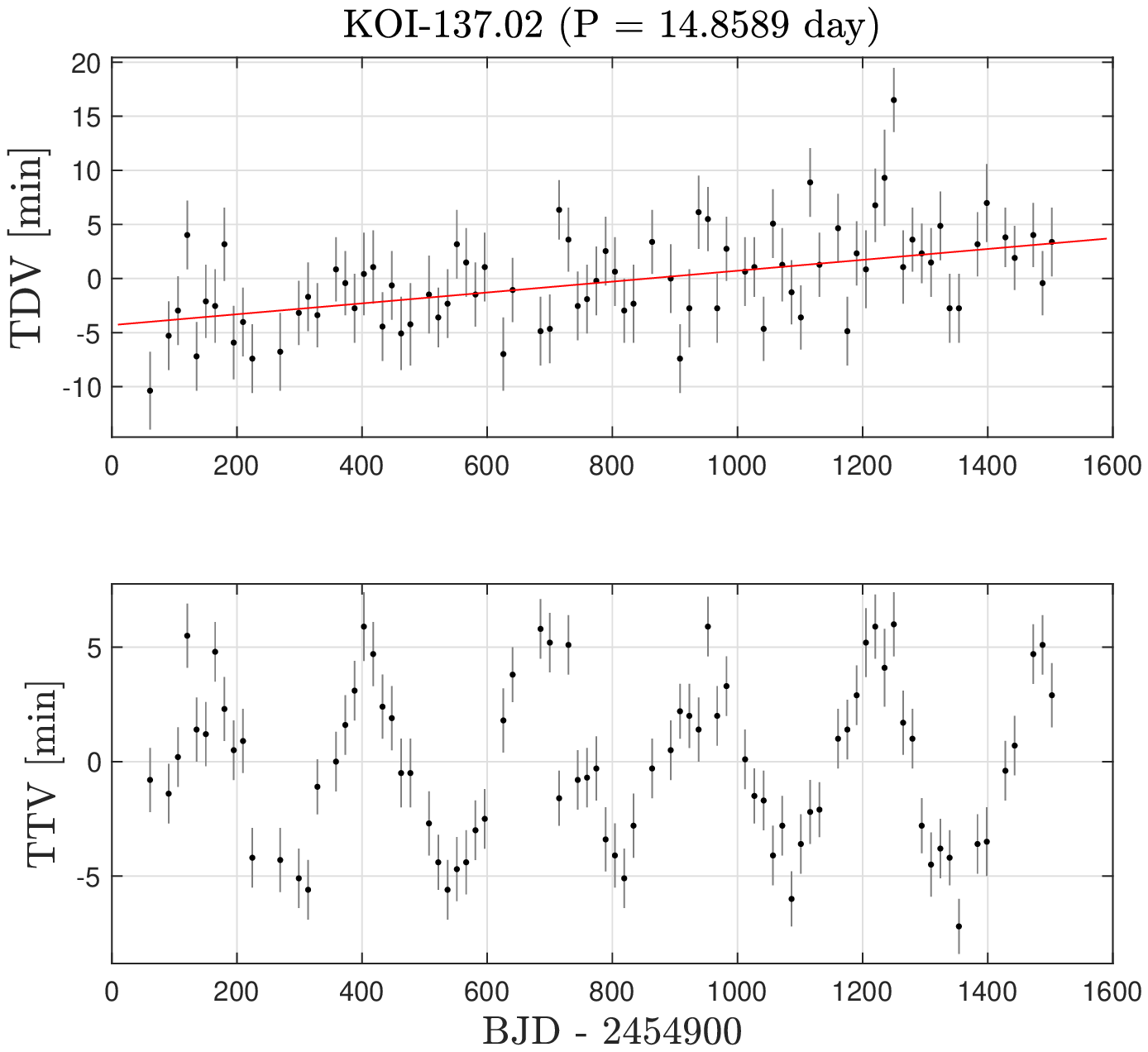}}
\hspace{1mm}
{\includegraphics[width=0.46\textwidth]{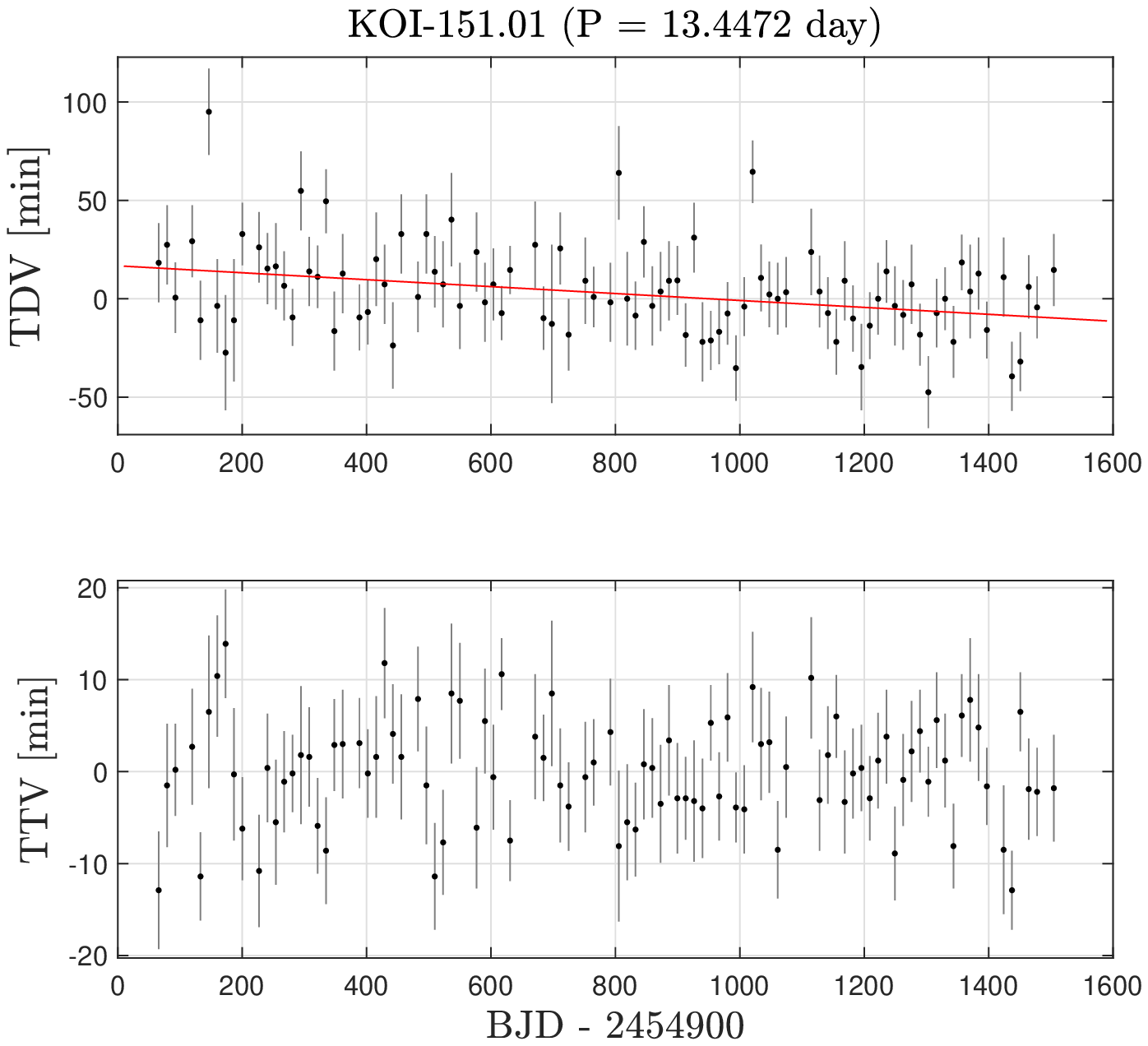}}\\ \vspace{5mm}
{\includegraphics[width=0.46\textwidth]{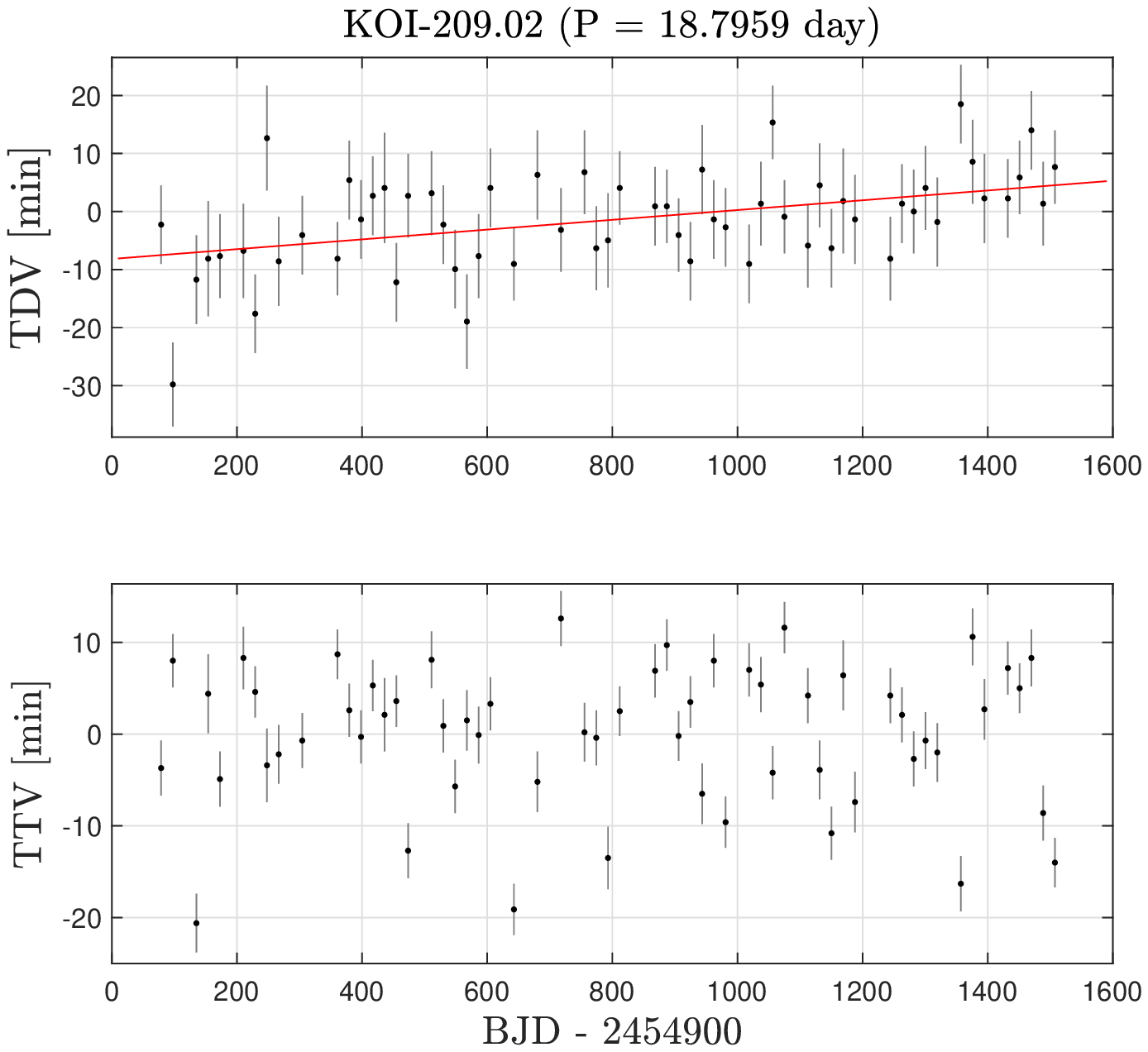}}
\hspace{1mm}
{\includegraphics[width=0.46\textwidth]{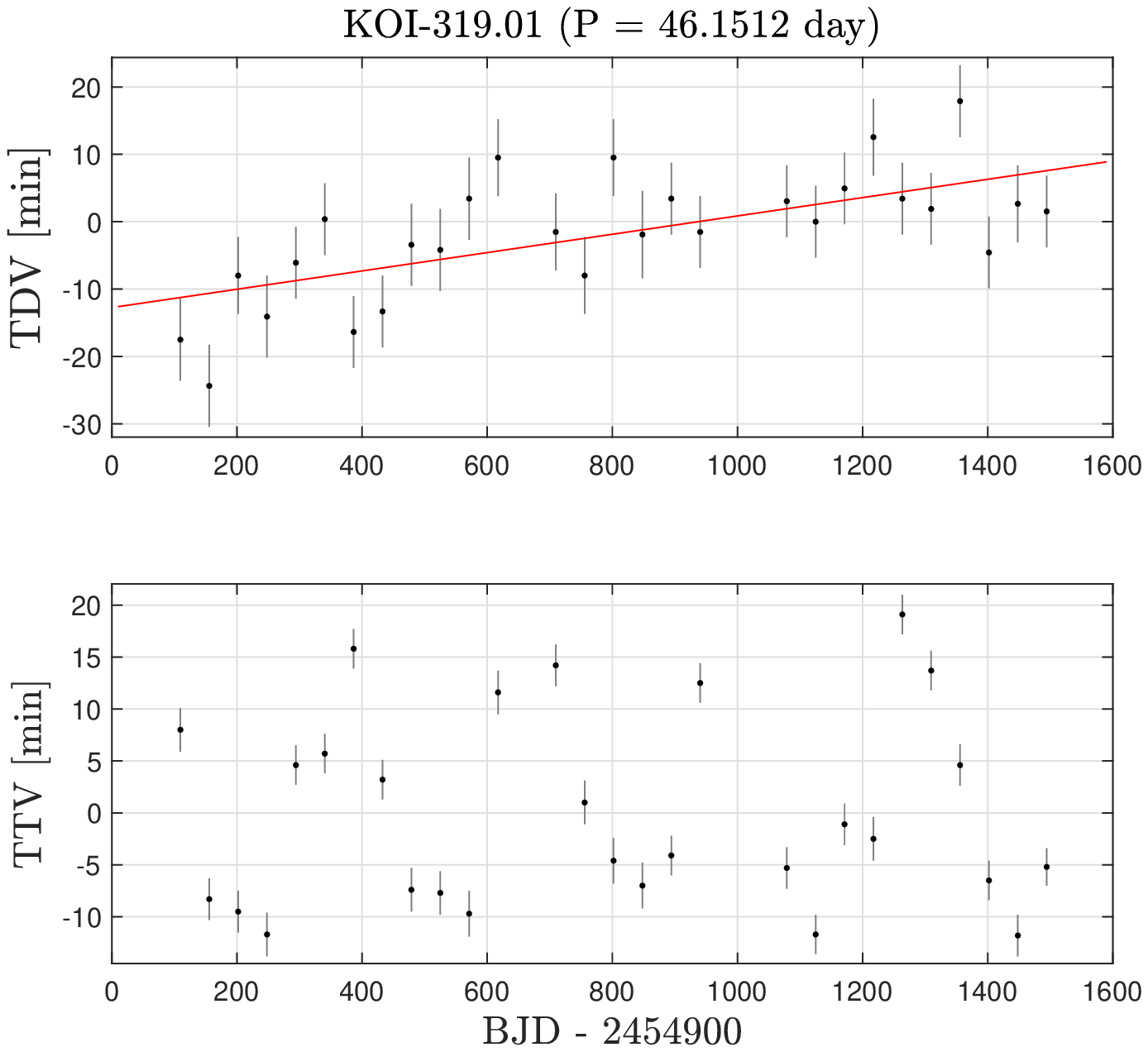}}
\caption{TDV and TTV plots for planets with significant long-term slope of transit duration (see Table~\ref{table:Significant}). Data taken from \citet{holczer16} catalogue. Red line is the fitted slope.}
\label{figure: TDV signif 1}
\end{figure*}

\begin{figure*}
\centering
{\includegraphics[width=0.46\textwidth]{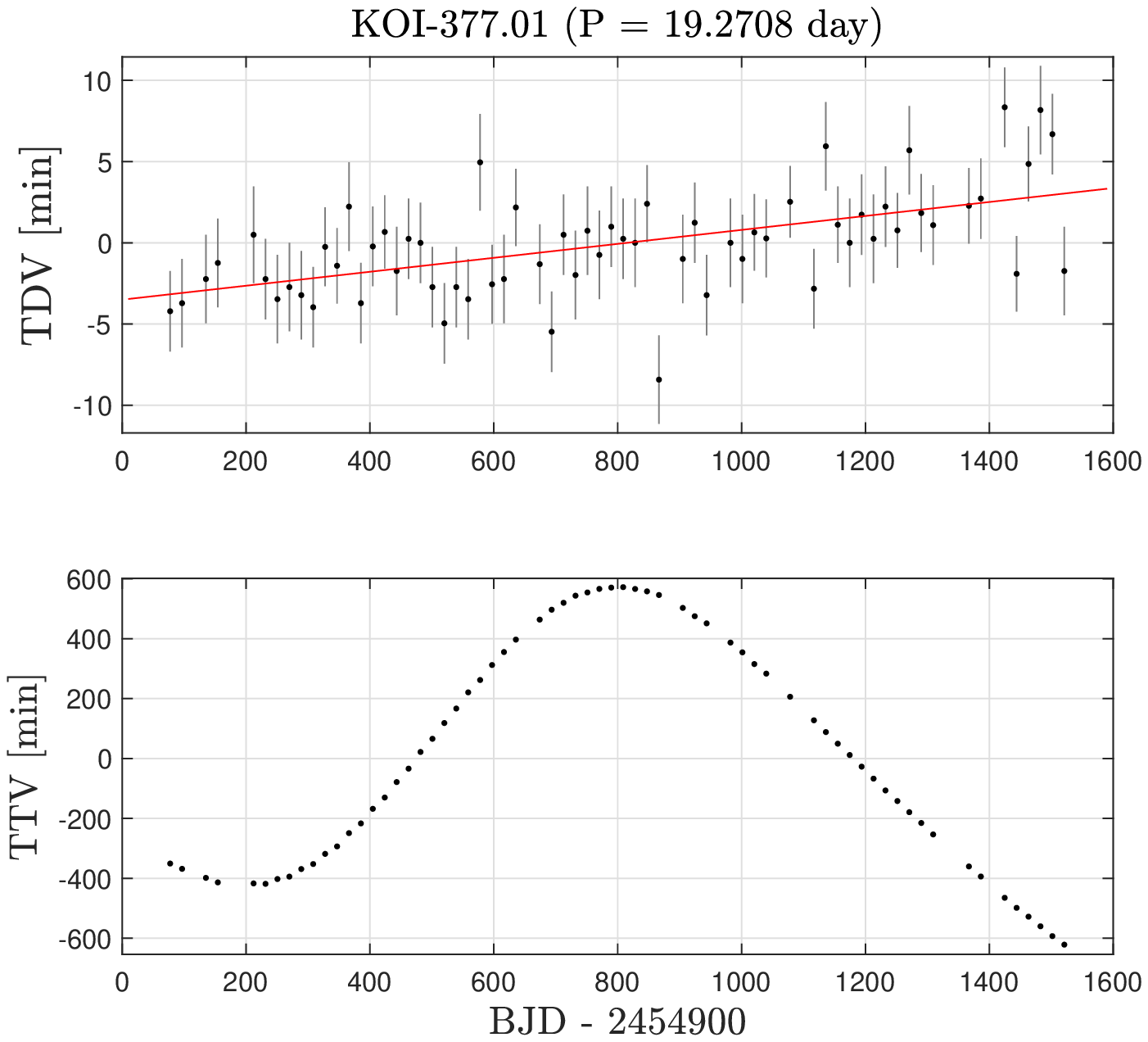}}
\hspace{1mm}
{\includegraphics[width=0.46\textwidth]{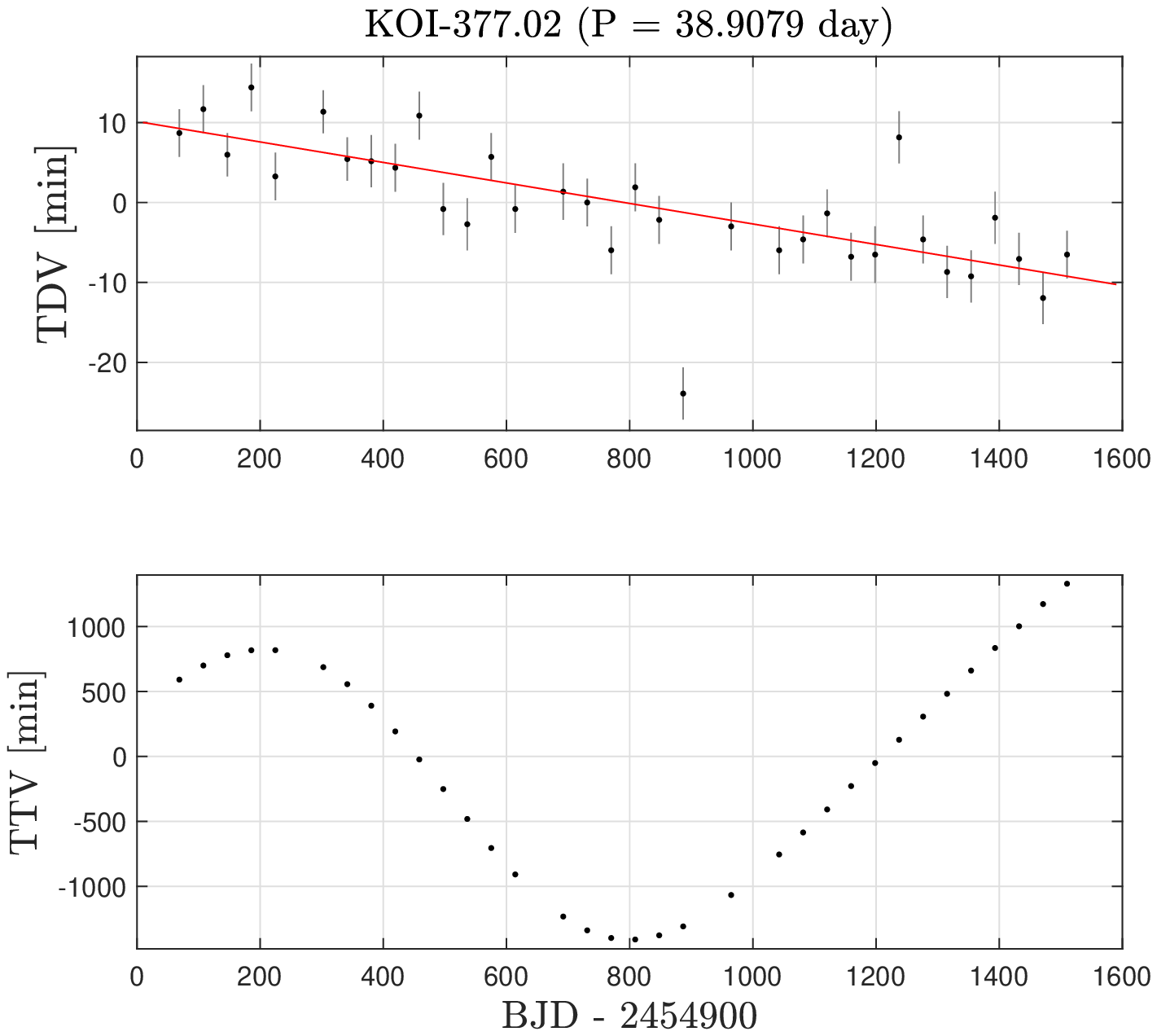}}\\ \vspace{5mm}
{\includegraphics[width=0.46\textwidth]{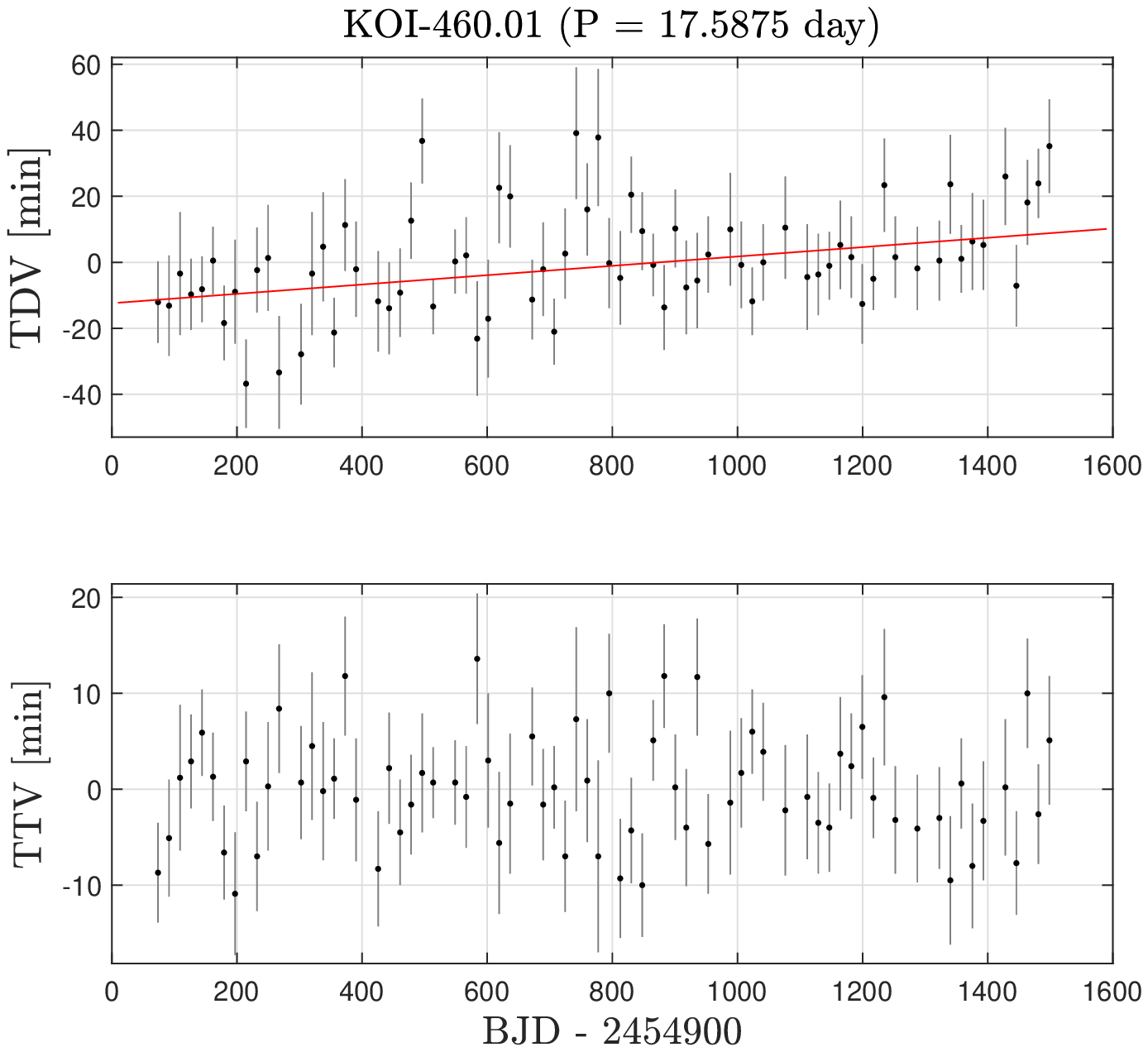}}
\hspace{1mm}
{\includegraphics[width=0.46\textwidth]{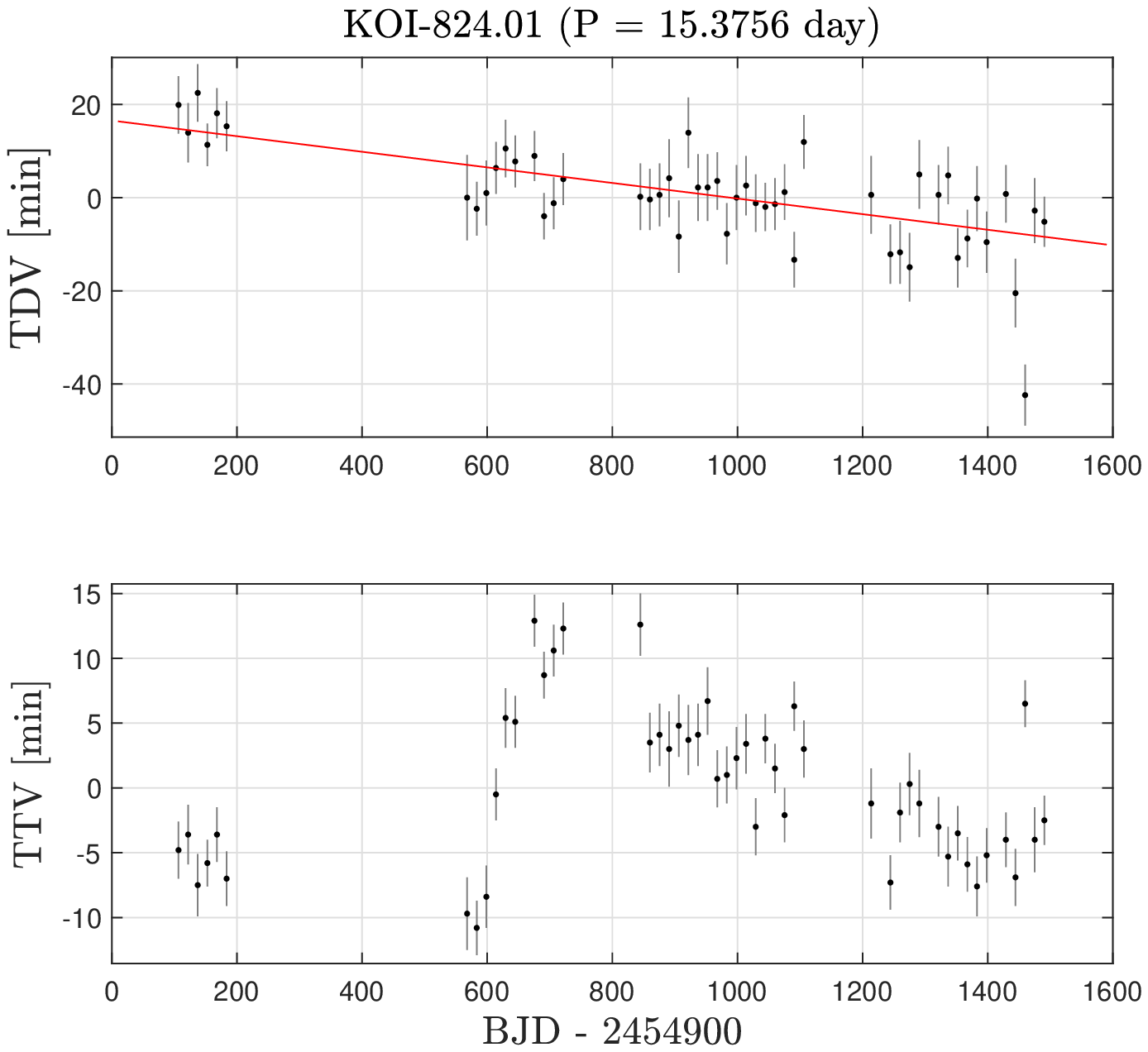}}\\ \vspace{5mm}
{\includegraphics[width=0.46\textwidth]{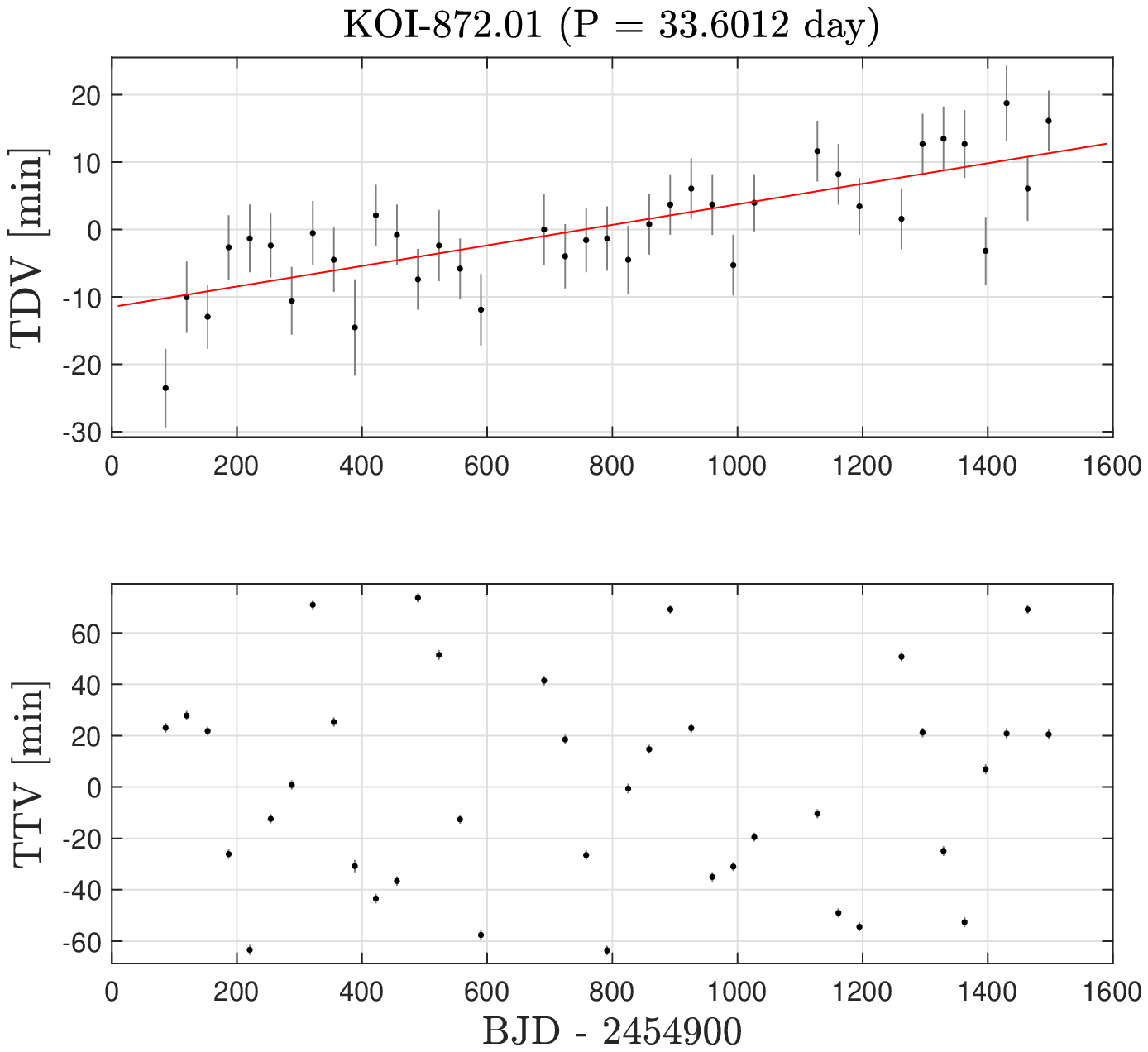}}
\hspace{1mm}
{\includegraphics[width=0.46\textwidth]{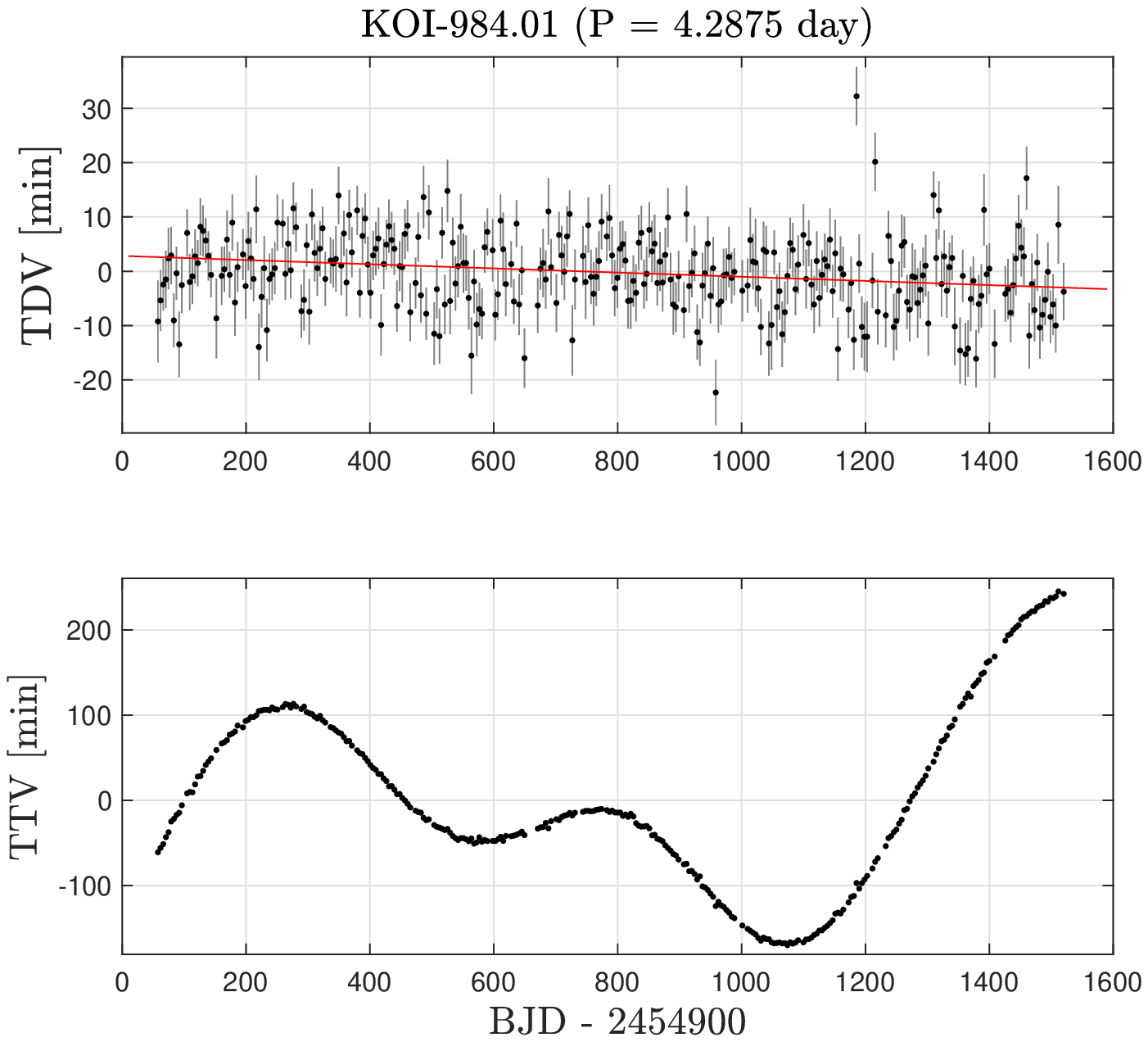}}
\caption{Additional plots for planets with significant long-term slope of transit duration. See Fig.~\ref{figure: TDV signif 1}.}
\label{figure: TDV signif 2}
\end{figure*}


\begin{figure*}
\centering
{\includegraphics[width=0.46\textwidth]{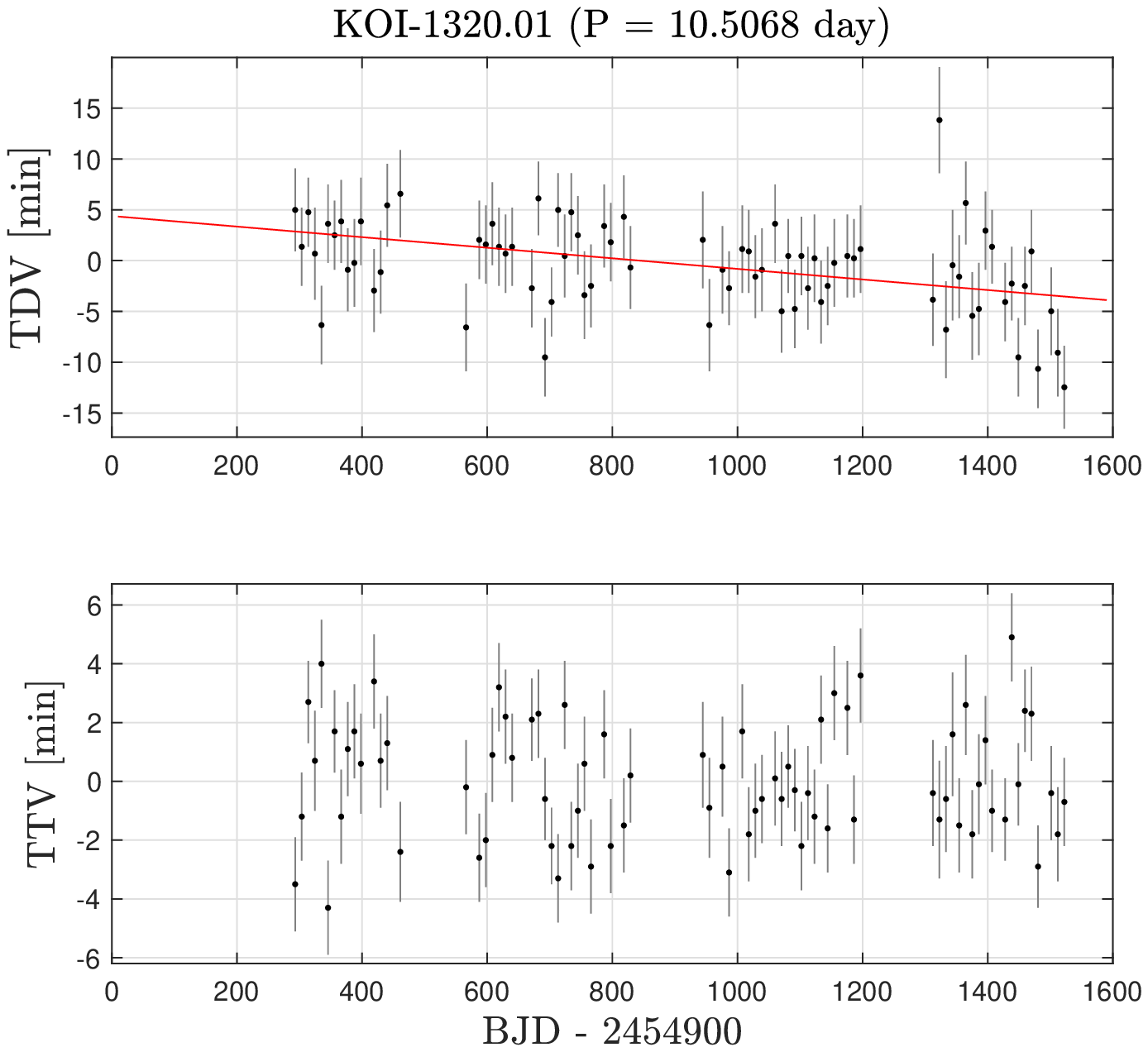}} \hspace{1mm}
{\includegraphics[width=0.46\textwidth]{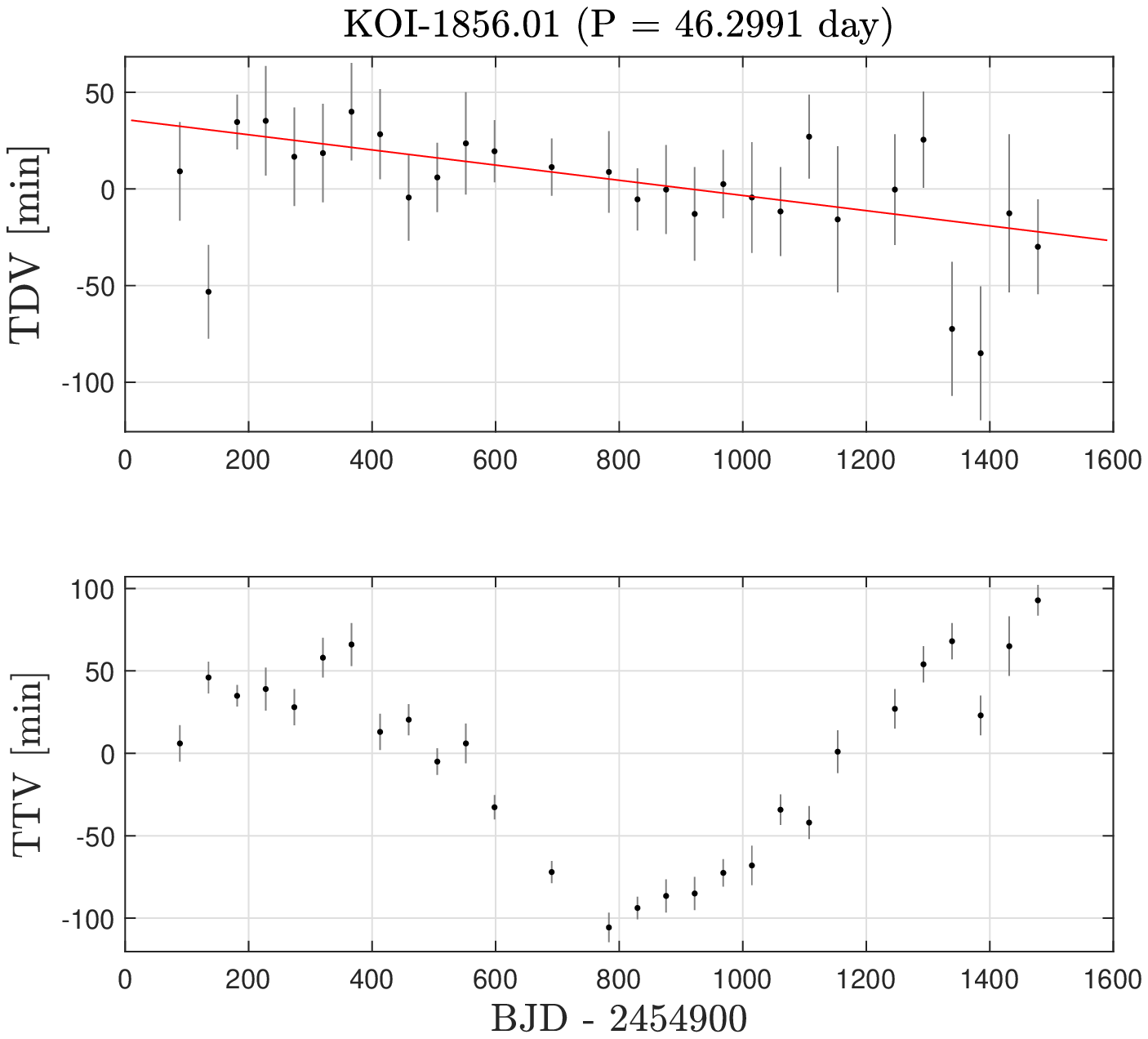}}\\ \vspace{5mm}
{\includegraphics[width=0.46\textwidth]{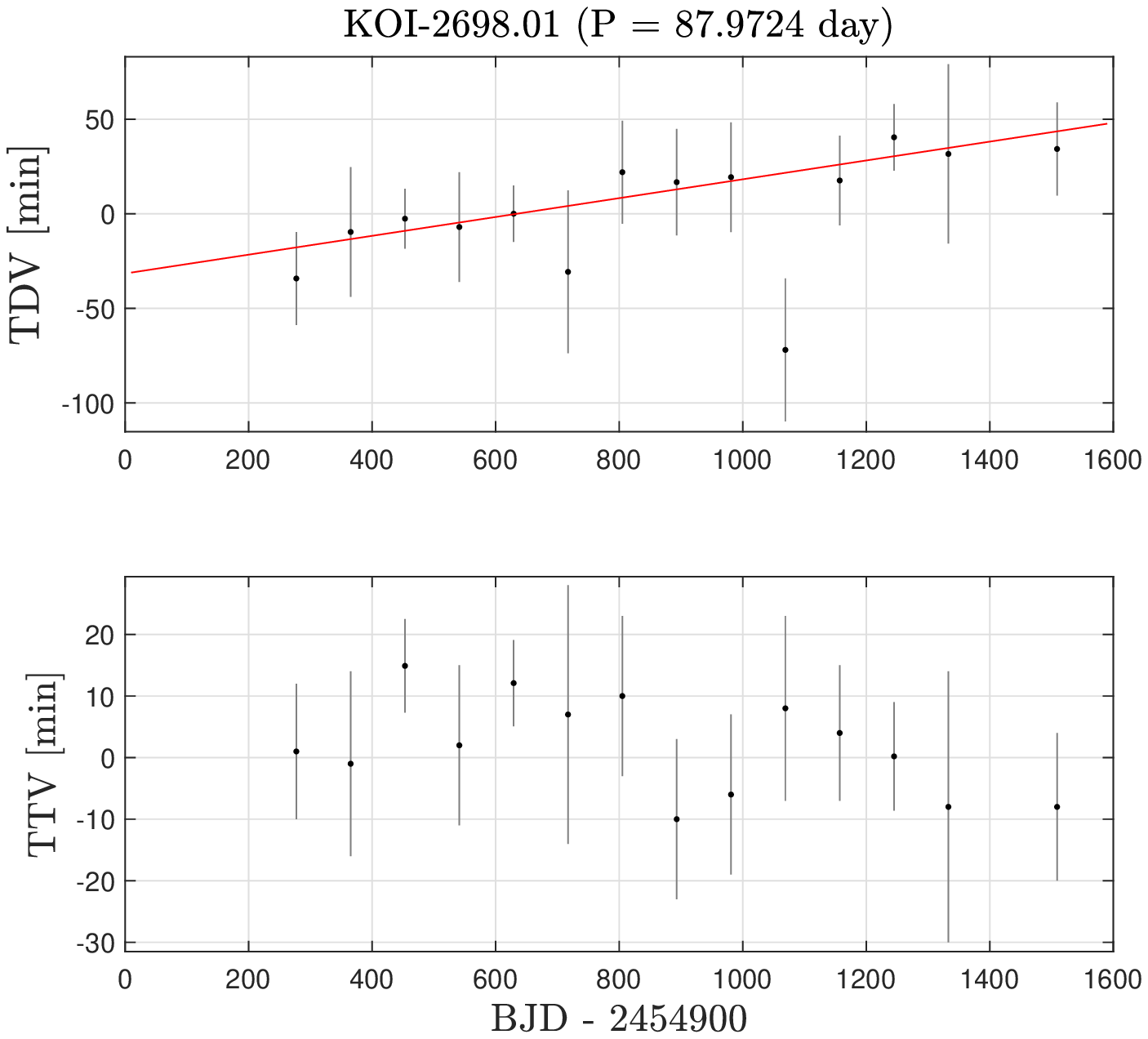}}\hspace{1mm}
{\includegraphics[width=0.46\textwidth]{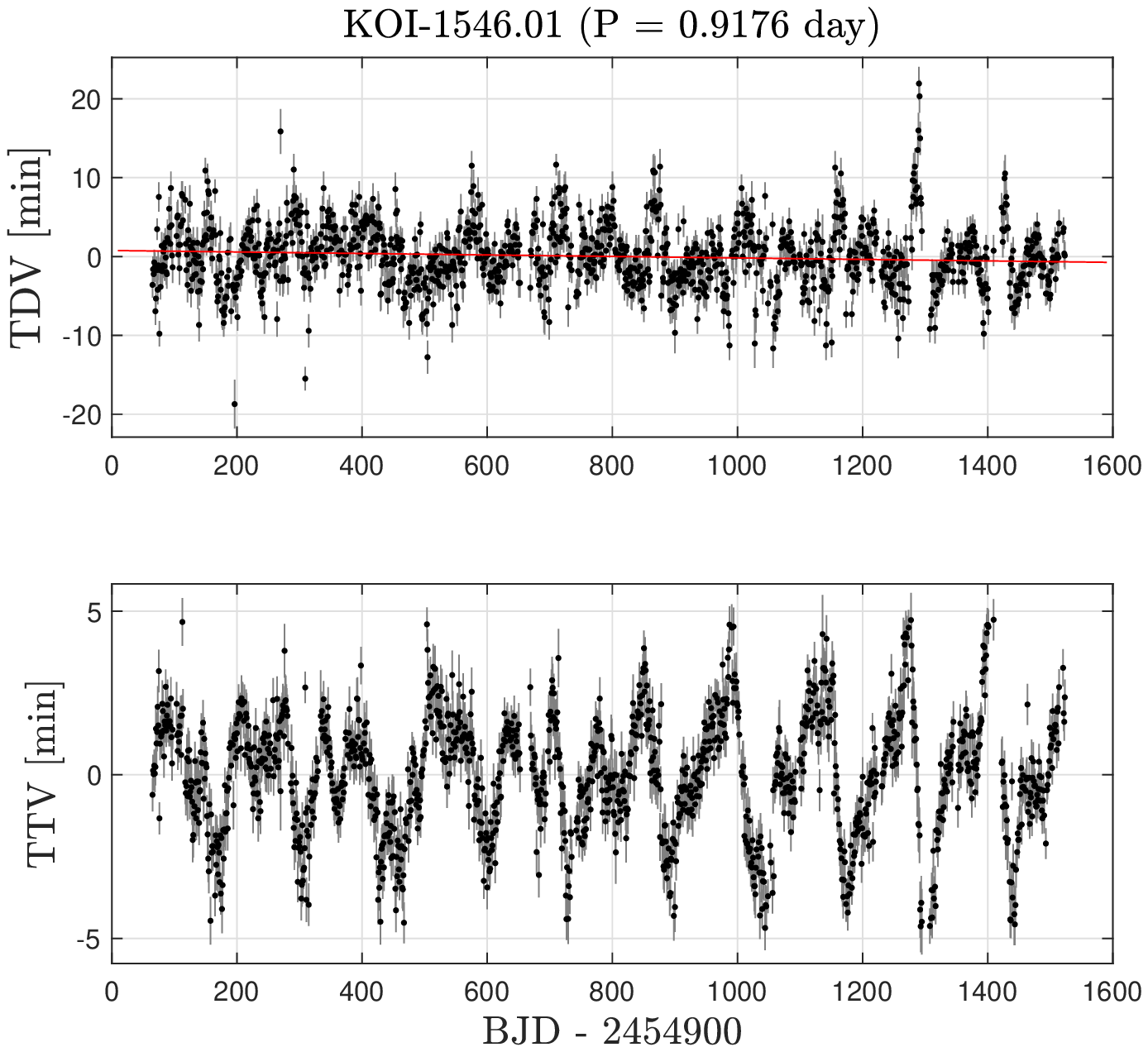}}
\caption{Additional plots for planets with significant long-term slope of transit duration. See Fig.~\ref{figure: TDV signif 1}.
KOI-1546.01 was rejected from the sample due to its $\mathcal{A}$ score (see text).}
\label{figure: TDV signif 3}
\end{figure*}
 
\begin{figure*}
\centering
{\includegraphics[width=0.46\textwidth]{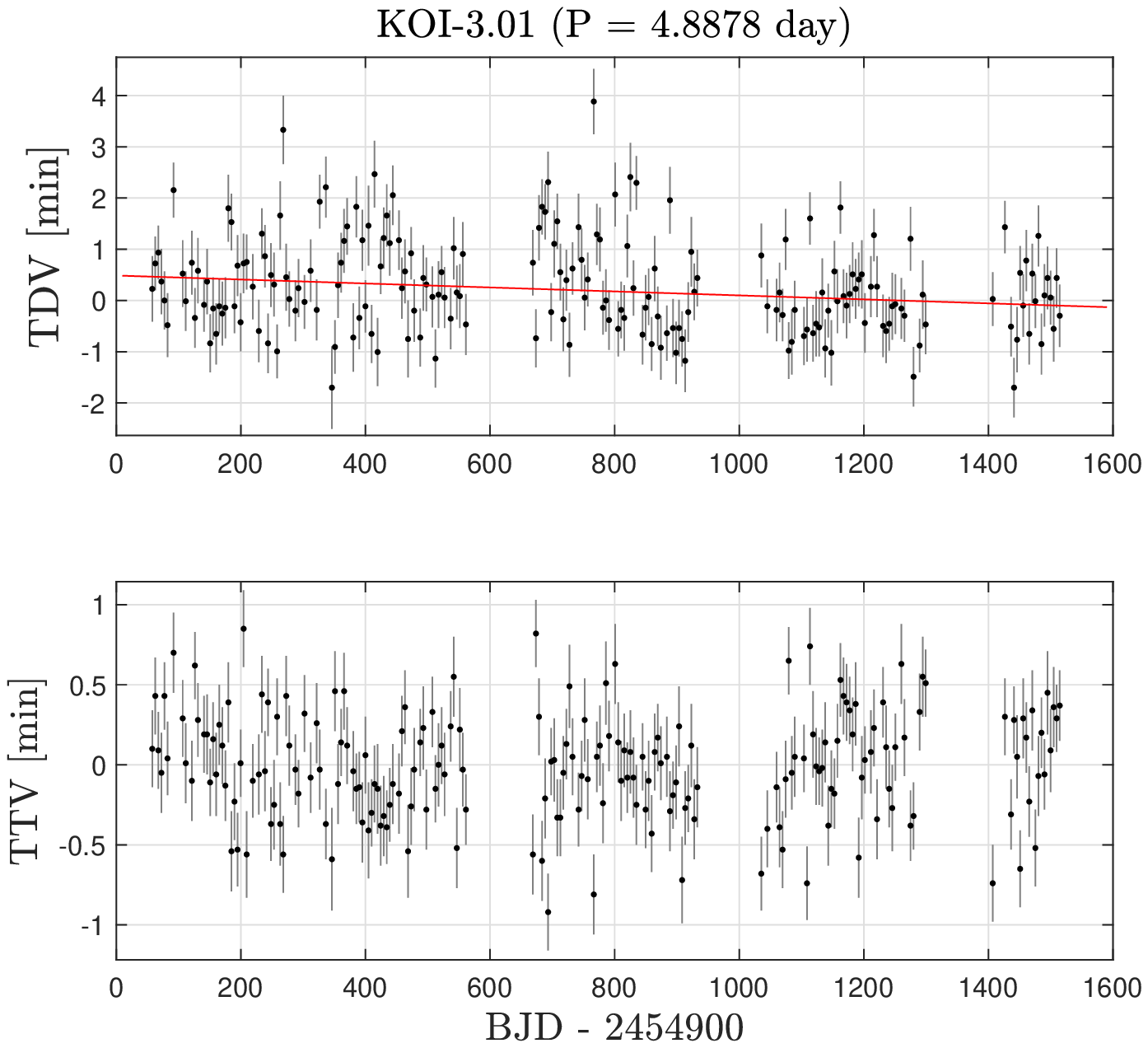}} \hspace{1mm}
{\includegraphics[width=0.46\textwidth]{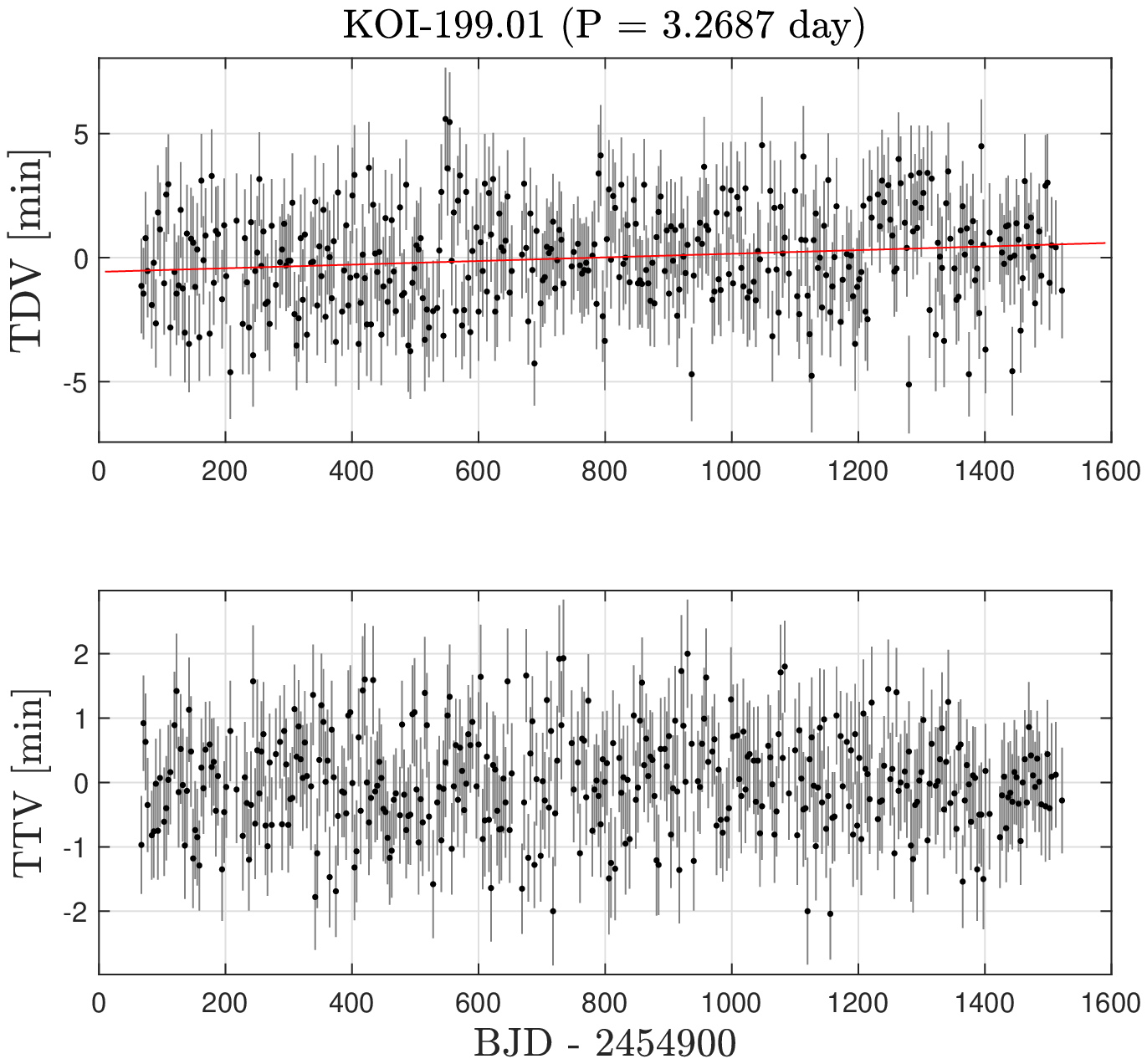}}\\ \vspace{5mm}
{\includegraphics[width=0.46\textwidth]{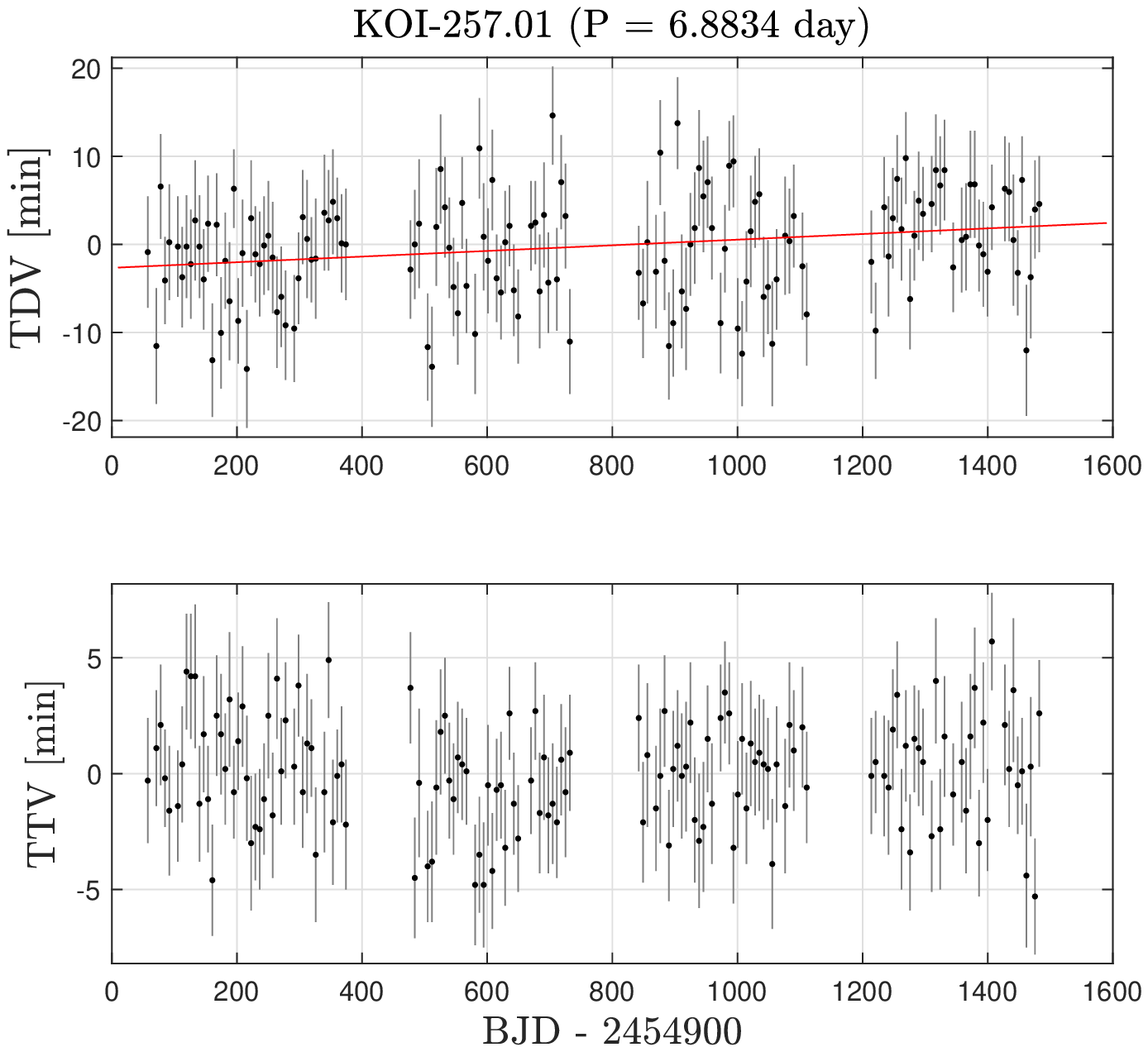}} \hspace{1mm}
{\includegraphics[width=0.46\textwidth]{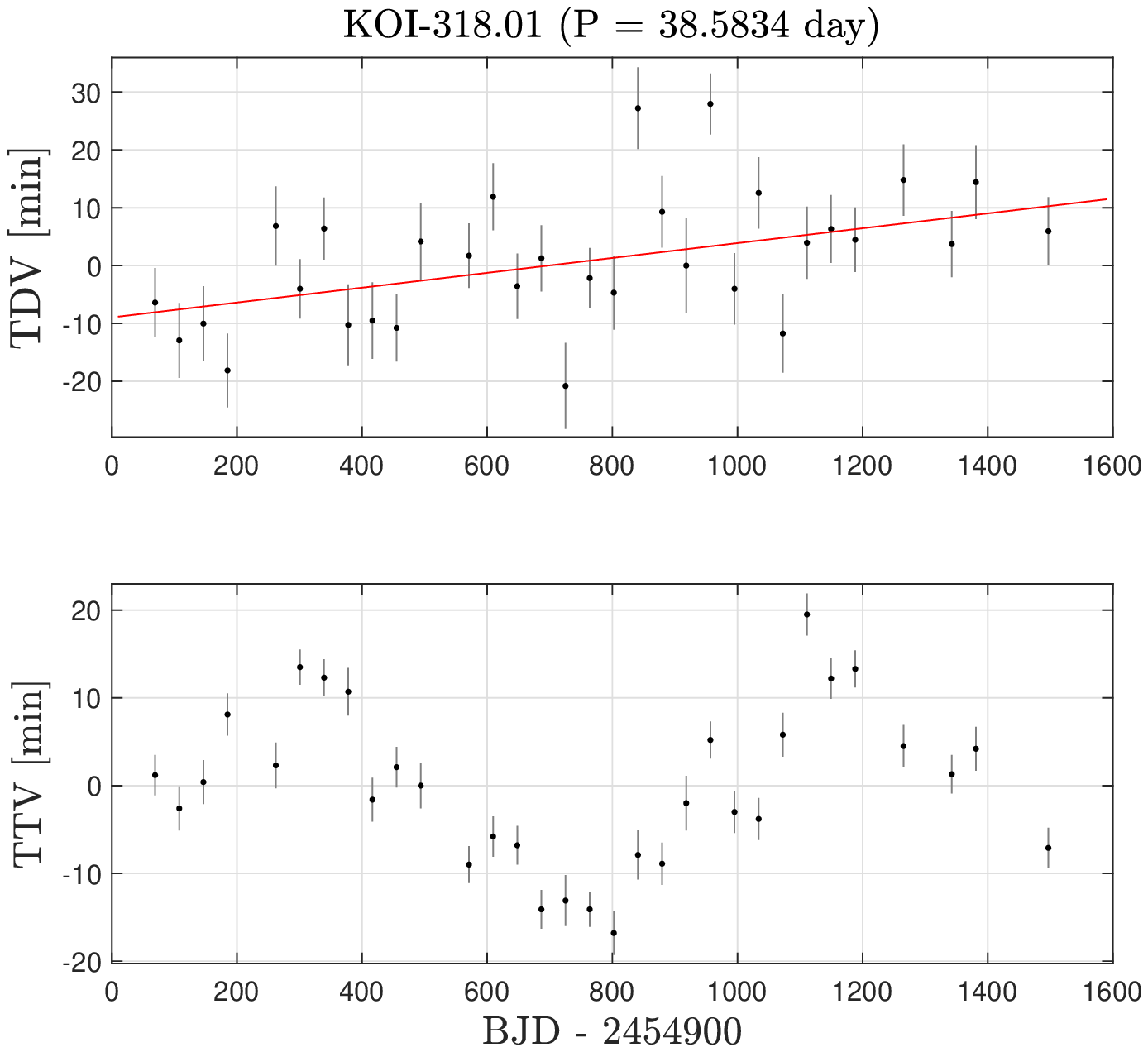}}\\ \vspace{5mm}
{\includegraphics[width=0.46\textwidth]{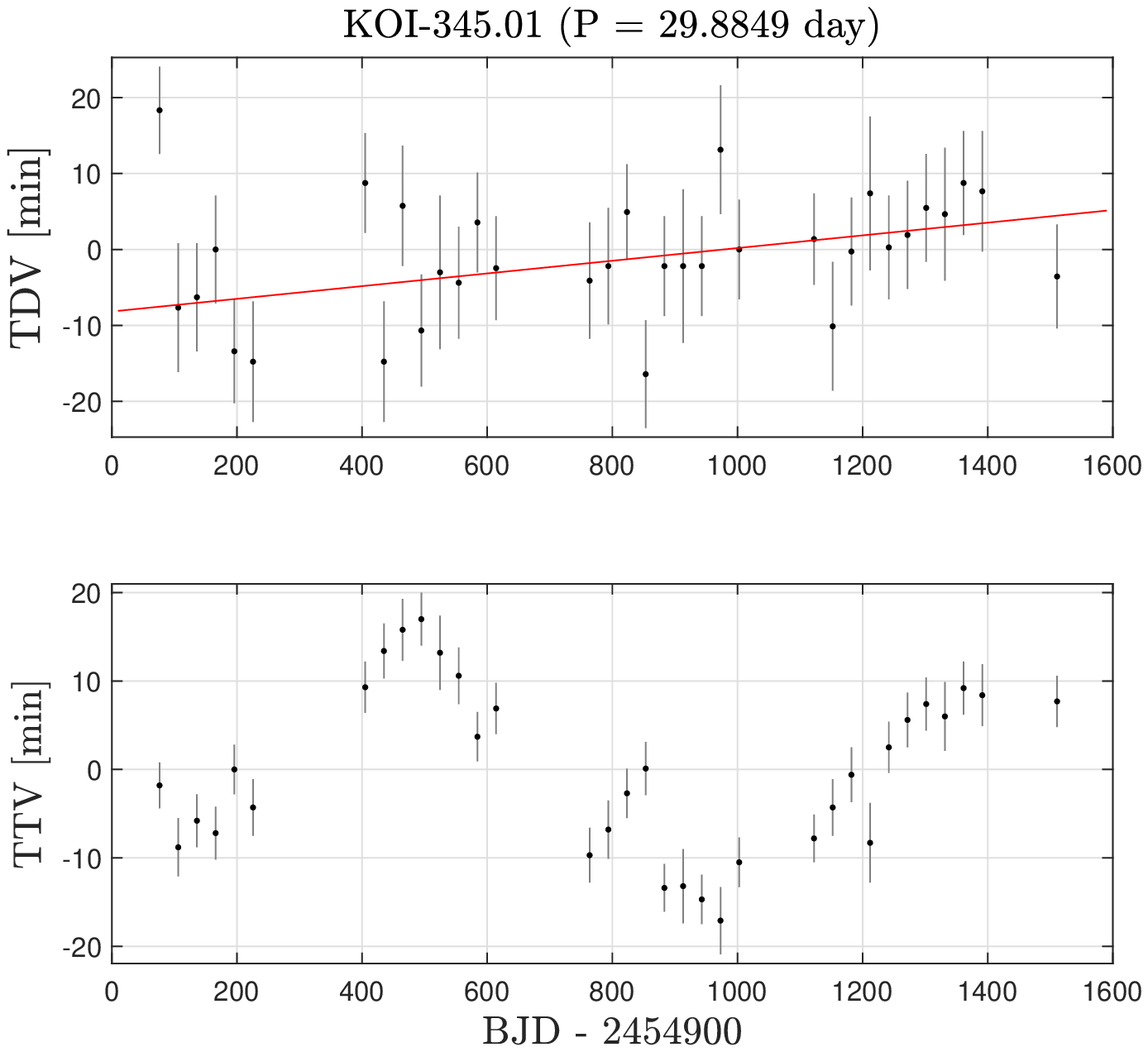}} \hspace{1mm}
{\includegraphics[width=0.46\textwidth]{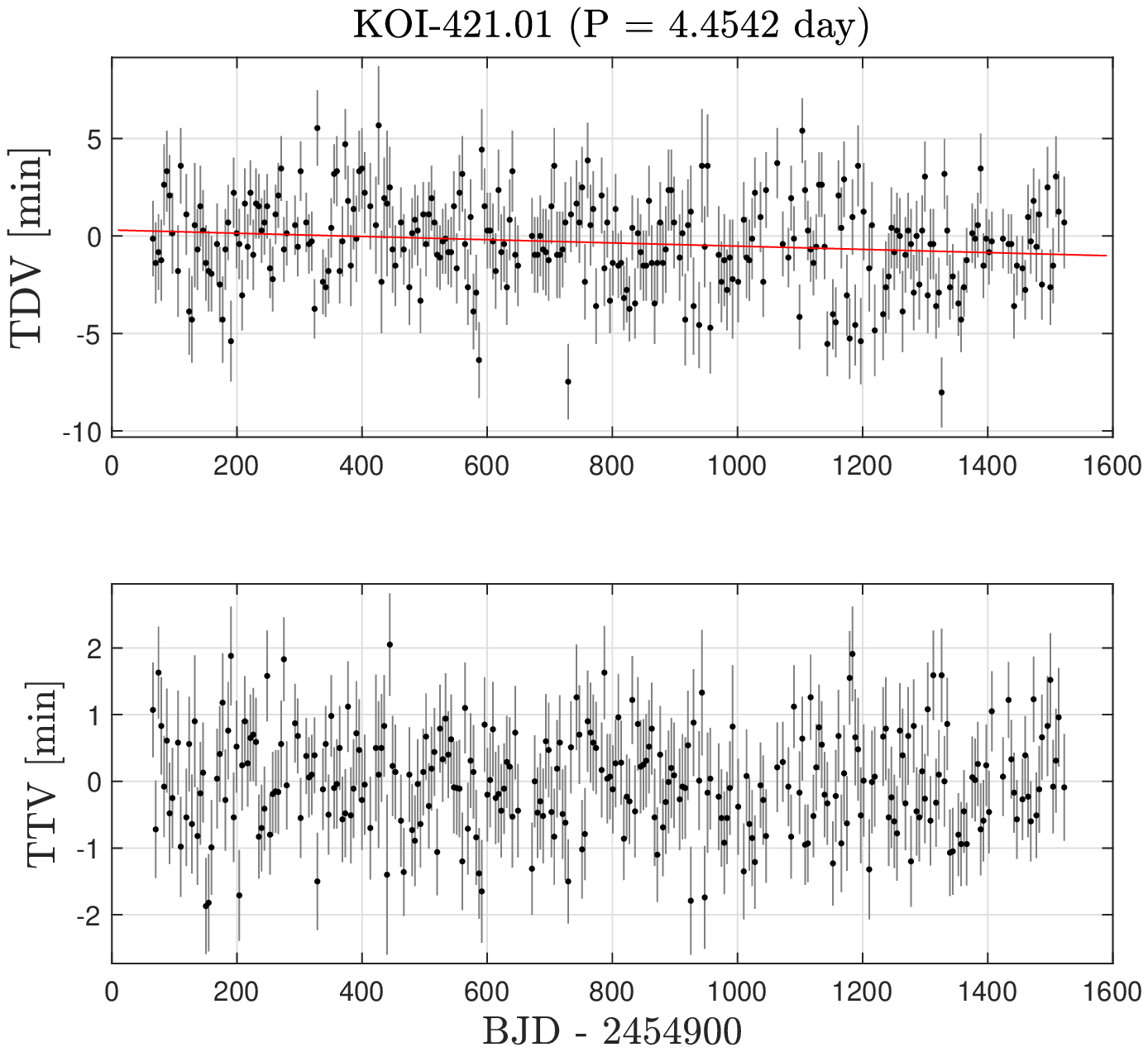}}
\caption{TDV and TTV plots for planets with long-term slope of transit duration of intermediate significance 
(see Table~\ref{table:Significant}). Details as in Fig.~\ref{figure: TDV signif 1}.}
\label{figure: TDV intermd 1}
\end{figure*}

\begin{figure*}
\centering
{\includegraphics[width=0.46\textwidth]{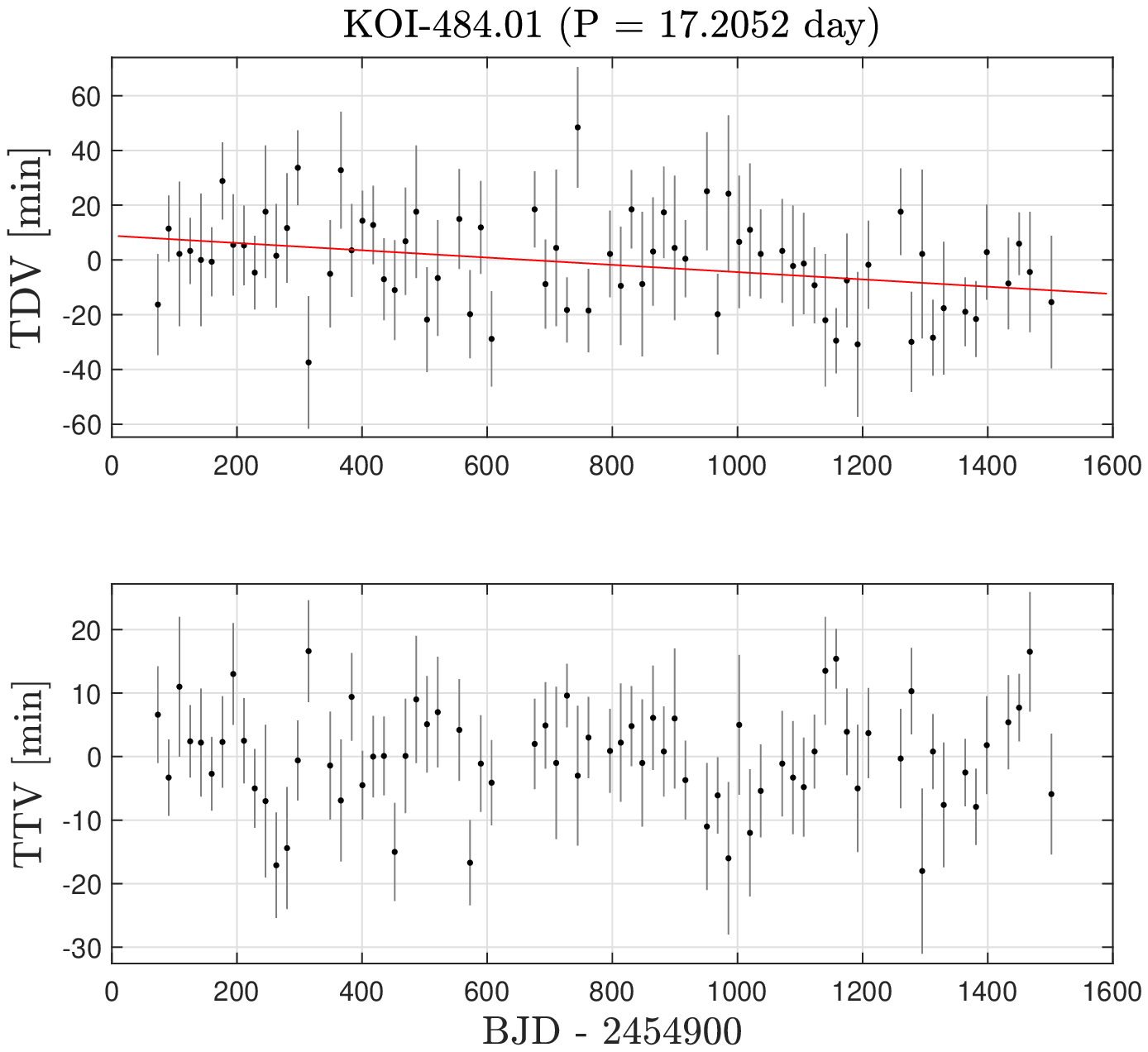}}\hspace{1mm}
{\includegraphics[width=0.46\textwidth]{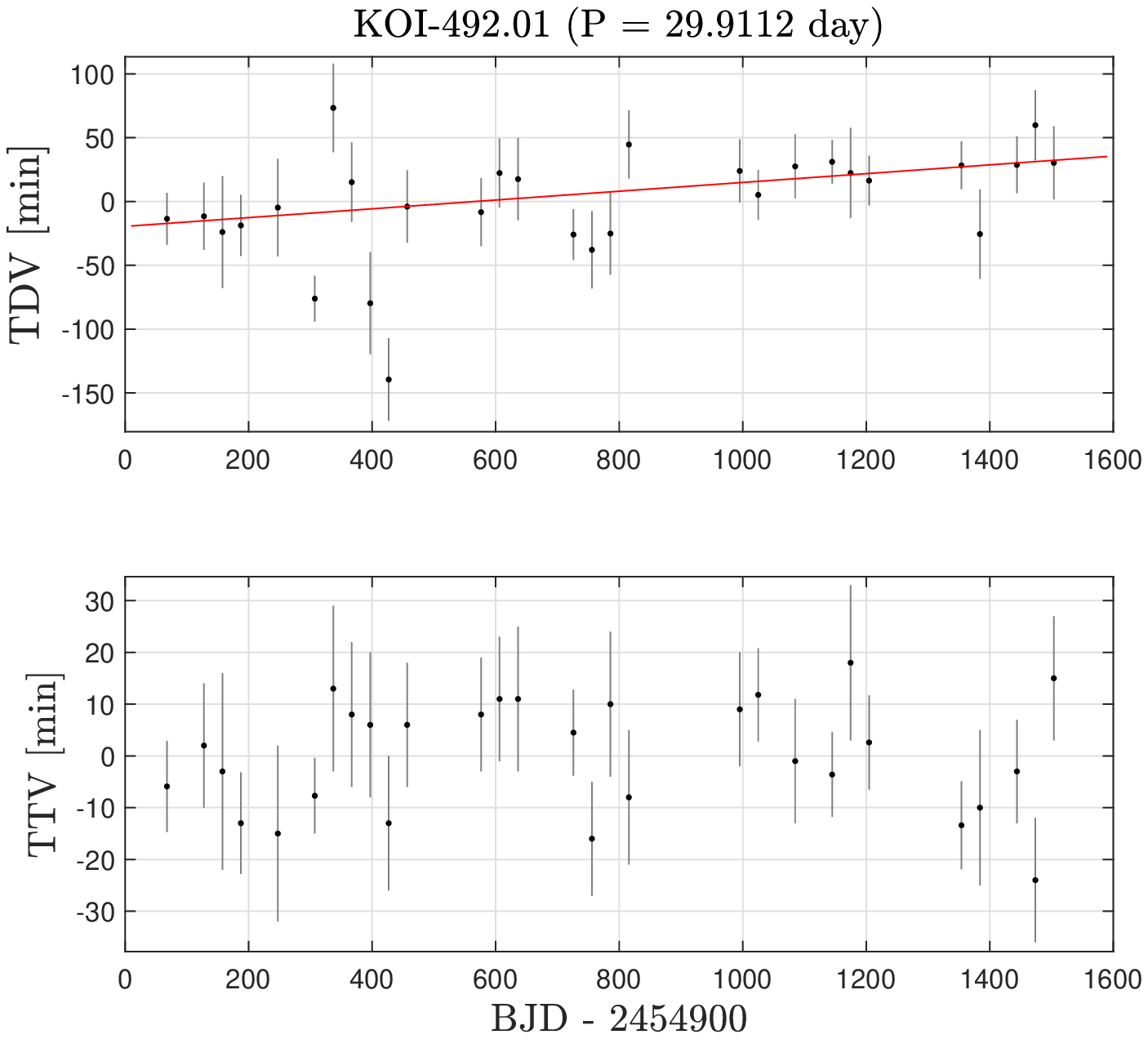}}\\ \vspace{5mm}
{\includegraphics[width=0.46\textwidth]{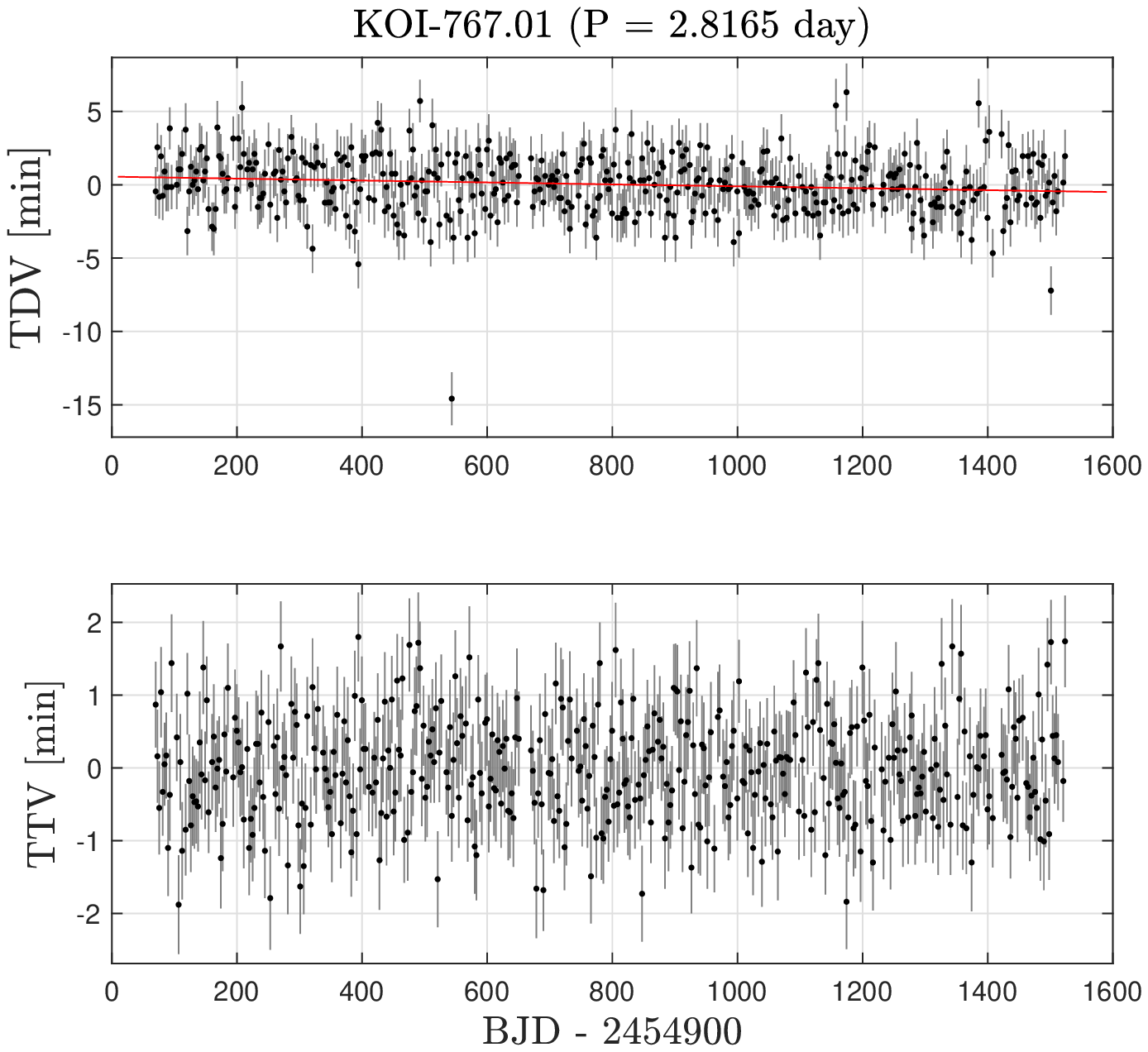}} \hspace{1mm} 
{\includegraphics[width=0.46\textwidth]{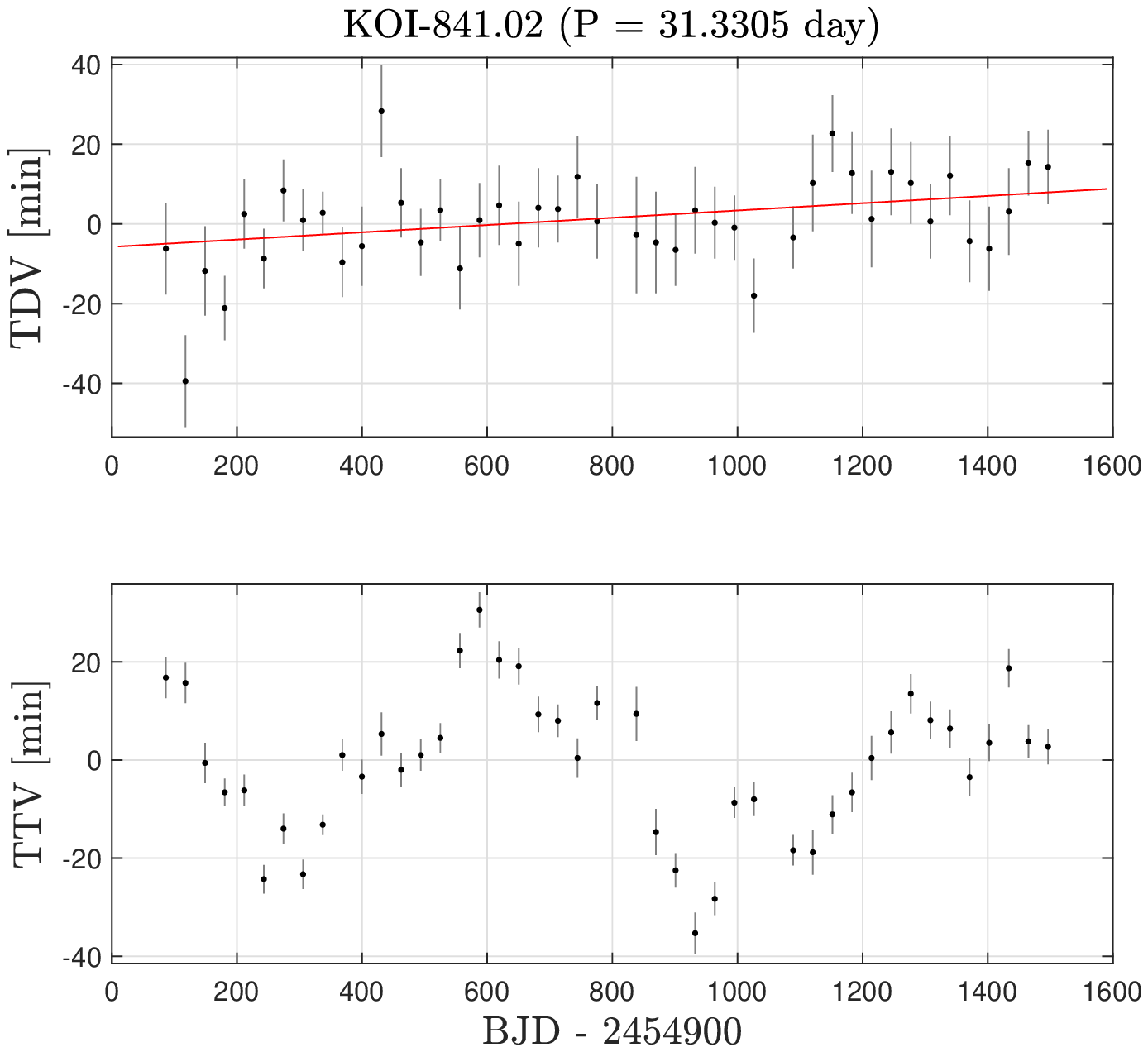}} \\ \vspace{5mm}
{\includegraphics[width=0.46\textwidth]{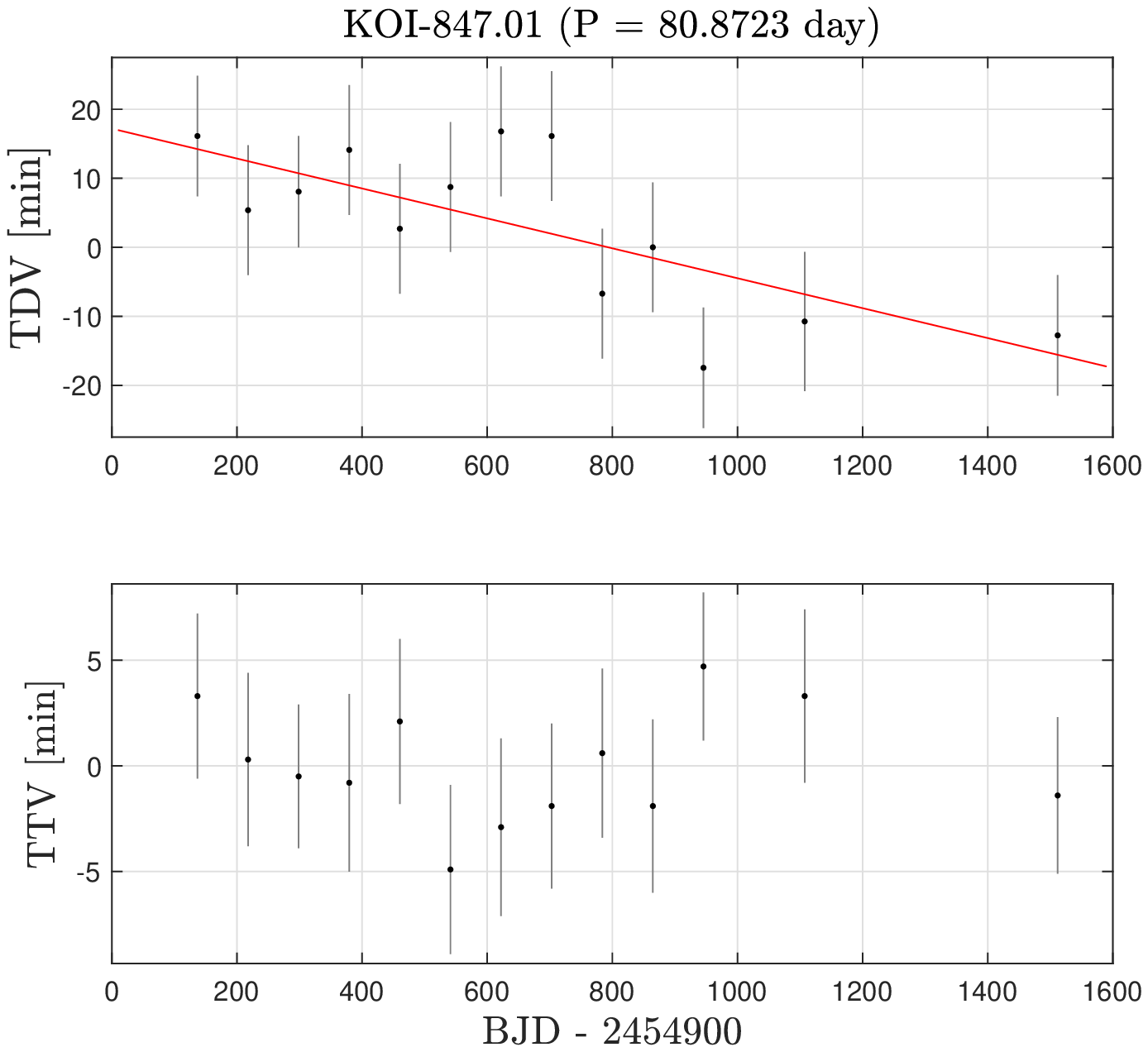}} \hspace{1mm} 
{\includegraphics[width=0.46\textwidth]{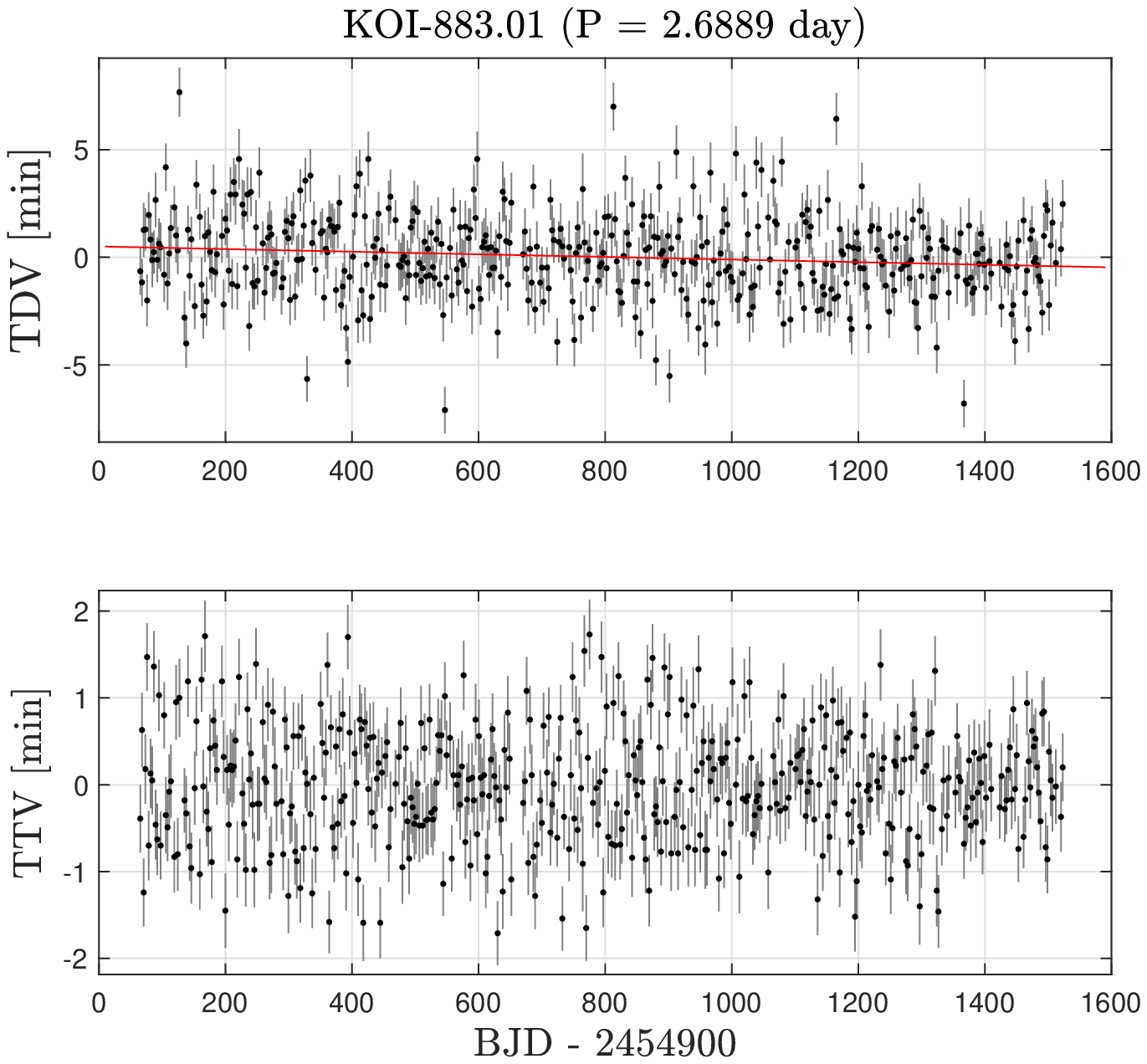}} 
\caption{Additional plots for planets with long-term slope of transit duration of intermediate significance (see Fig.~\ref{figure: TDV intermd 1}).}
\label{figure: TDV intermd 2}
\end{figure*}

\begin{figure*}
\centering
{\includegraphics[width=0.46\textwidth]{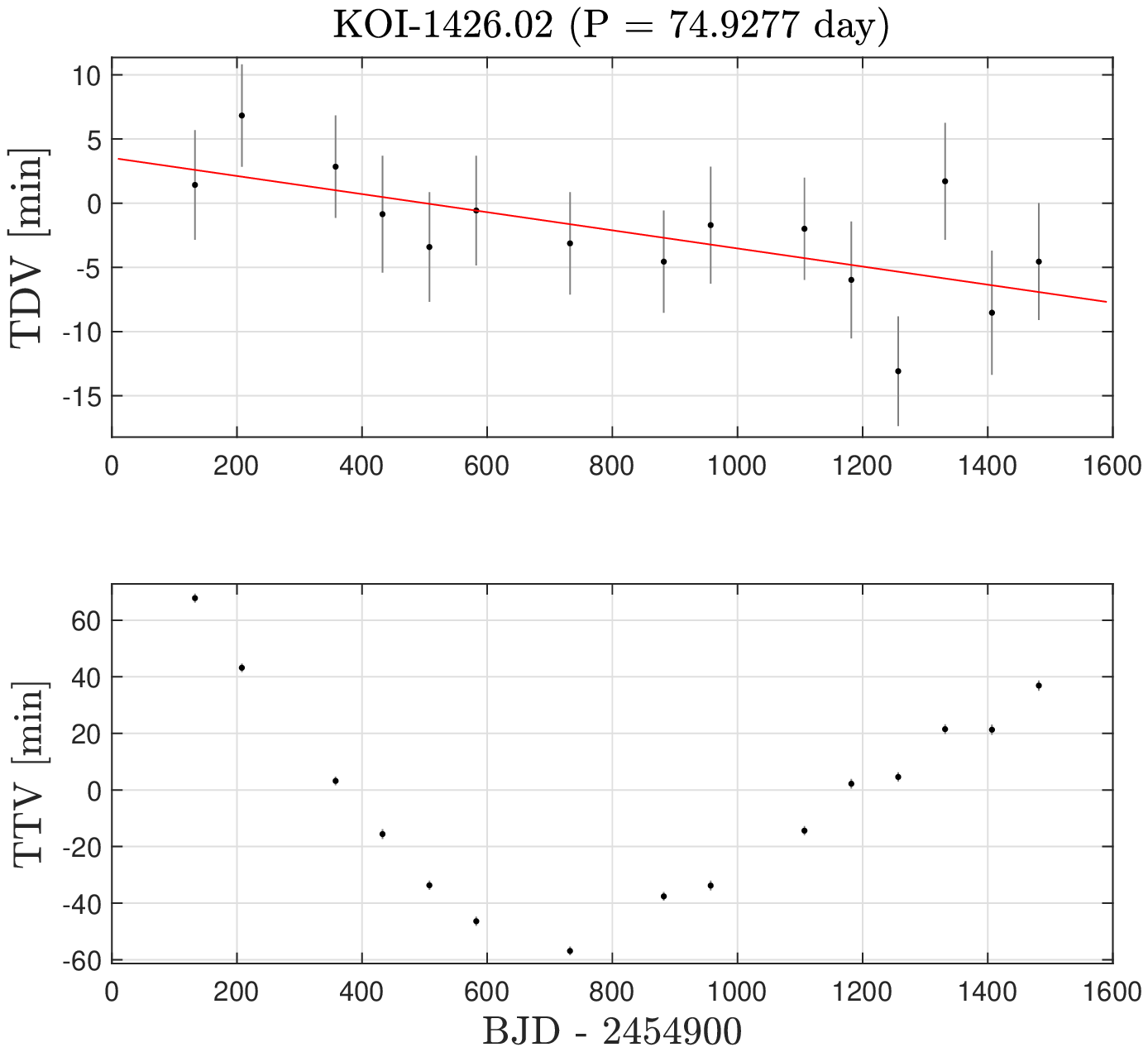}} \hspace{1mm}
{\includegraphics[width=0.46\textwidth]{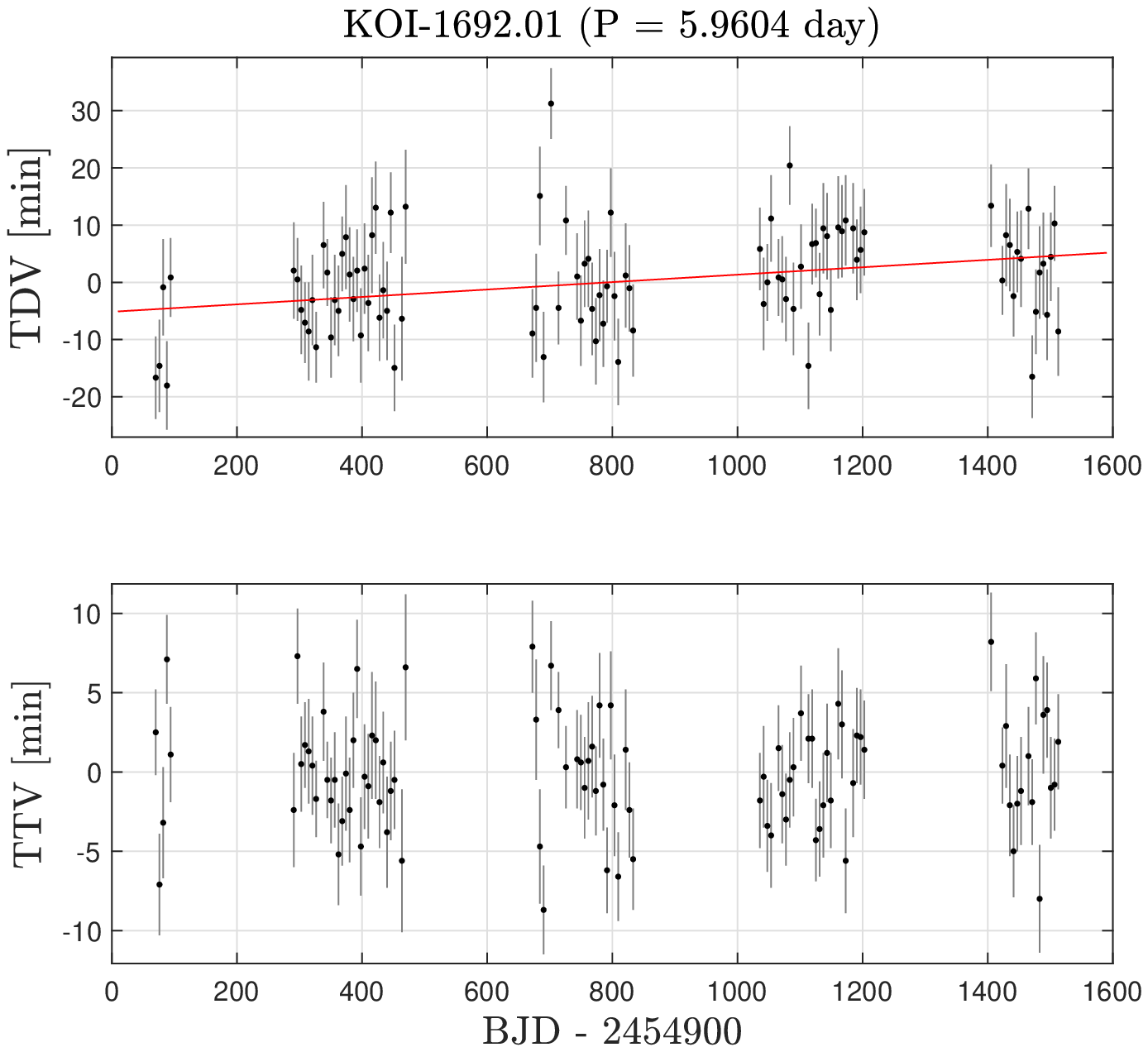}}\\ \vspace{5mm}
{\includegraphics[width=0.46\textwidth]{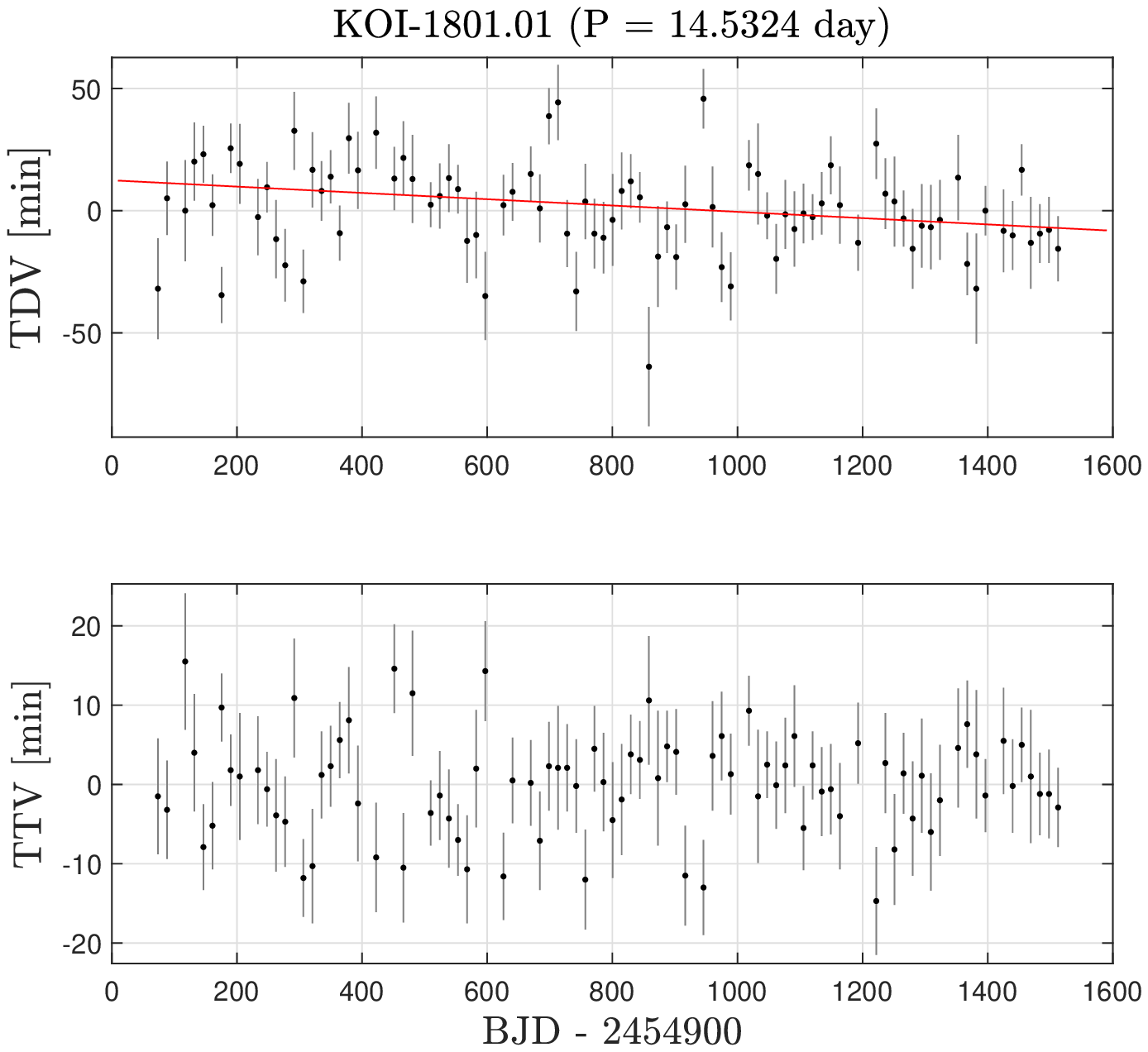}}\hspace{1mm}
{\includegraphics[width=0.46\textwidth]{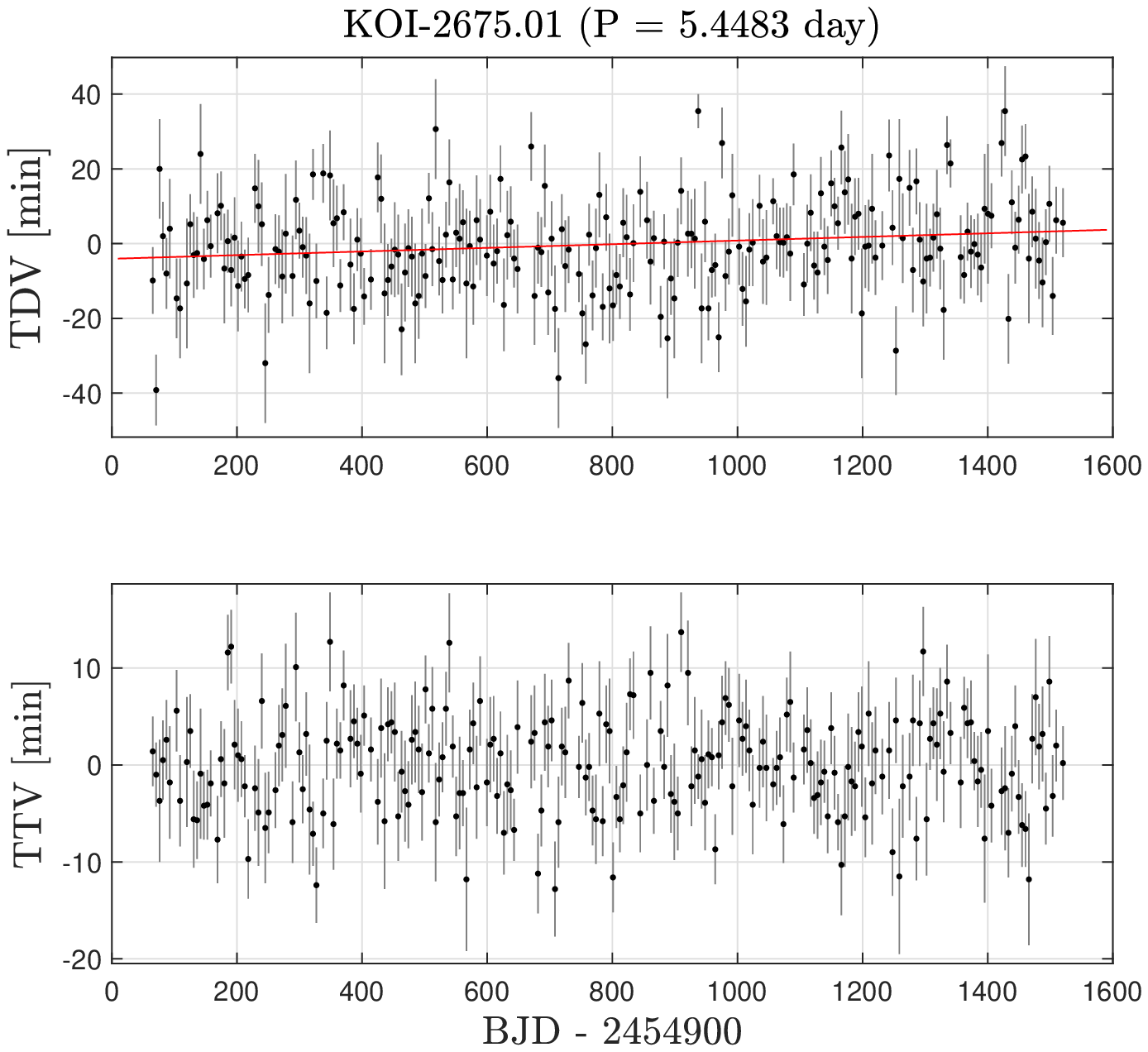}}
\caption{Additional plots for planets with long-term slope of transit duration of intermediate significance (see Fig.~\ref{figure: TDV intermd 1}).}
\label{figure: TDV intermd 3}
\end{figure*}

\newpage
\bsp	
\label{lastpage}
\end{document}